\documentclass[fleqn,usenatbib]{mnras}

\usepackage{newtxtext,newtxmath}

\usepackage[T1]{fontenc}

\DeclareRobustCommand{\VAN}[3]{#2}
\let\VANthebibliography\thebibliography
\def\thebibliography{\DeclareRobustCommand{\VAN}[3]{##3}\VANthebibliography}

\usepackage{graphicx}	
\usepackage{amsmath}	

\usepackage{amssymb}	

\newcommand{\kms}{\mbox{$\>{\rm km\, s^{-1}}$}}
\newcommand{\masyr}{\mbox{$\>{\rm mas\, yr^{-1}}$}}
\newcommand{\kpc}{\mbox{$\>{\rm kpc}$}}
\newcommand{\pc}{\mbox{$\>{\rm pc}$}}
\newcommand{\Gyr}{\mbox{$\>{\rm Gyr}$}}

\newcommand{\Msun}{\>{\rm M_{\odot}}}

\newcommand\degrees{^\circ}
\newcommand{\avg}[1]{\mbox{$\left<{#1}\right>$}}

\newcommand{\feh}{\mbox{$\rm [Fe/H]$}}

\newcommand{\mags}{\mbox{$\>{\rm mag}$}}

\newcommand{\gaia}{{\it Gaia}}

\newcommand{\pml}{\mbox{$\mu_l$}}
\newcommand{\apml}{\mbox{\avg{\pml}}}
\newcommand{\kpcmasyr}{\mbox{$\>{\rm kpc \cdot mas\, yr^{-1}}$}}

\def\etal{{et al.}}
\def\eg{{\it e.g.}}

\def\ie{{\it i.e.}}

\def\araa{ARAA}
\def\aj{AJ}
\def\apj{ApJ}
\def\apjl{ApJ}
\def\apjs{ApJS}
\def\aap{A\&A}
\def\mnras{MNRAS}
\def\pasa{Publ. Astron. Soc. Aust.} 
\def\nat{Nature}

\defcitealias{clarkson+18}{C18}

\title[MW Bulge Proper Motions]{Predicted Trends in Milky Way Bulge Proper Motion Rotation Curves: future Prospects for {\it HST} and LSST}

\author[Gough-Kelly et al.]{
Steven Gough-Kelly$^{1}$\thanks{E-mail: sgoughkelly@gmail.com},
Victor P. Debattista$^{1}$, William I. Clarkson$^{2}$,
\newauthor Oscar A. Gonzalez$^{3}$, Stuart R. Anderson$^{1}$, Mario Gennaro$^{4}$,
\newauthor Annalisa Calamida$^{4}$, Kailash C. Sahu$^{4}$
\\
$^{1}$Jeremiah Horrocks Institute, University of Central Lancashire, Preston PR1 2HE, UK\\
$^{2}$Department of Natural Sciences, University of Michigan-Dearborn, 4901 Evergreen Road, Dearborn, MI 48128, USA\\
$^{3}$UK Astronomy Technology Centre, Royal Observatory, Blackford Hill, Edinburgh EH9 3HJ, UK\\
$^{4}$Space Telescope Science Institute, 3700 San Martin Drive, Baltimore, MD 21218, USA
}

\date{Accepted XXX. Received YYY; in original form ZZZ}

\pubyear{2022}

\begin{document}
\label{firstpage}
\pagerange{\pageref{firstpage}--\pageref{lastpage}}
\maketitle

\begin{abstract}
We use an $N$-body+smoothed particle hydrodynamics simulation of an isolated barred galaxy to study the age dependence of bulge longitudinal proper motion (\pml) rotation curves. We show that close to the minor axis ($|l| \sim 0\degrees$) the relatively young stars rotate more rapidly than the old stars, as found by {\it Hubble Space Telescope} in the Milky Way's (MW's) bulge. This behaviour would be expected also if the MW were unbarred.  At larger $|l|$ a different behaviour emerges. Because younger stars trace a strong bar, their galactocentric radial motions dominate their \pml\ at $|l| \sim 6\degrees$, leading to a reversal in the sign of \apml. This results in a rotation curve with forbidden velocities (negative \apml\ at positive longitudes, and positive \apml\ at negative longitudes). The old stars, instead, trace a much weaker bar and thus their kinematics are more axisymmetric, resulting in no forbidden velocities. We develop metrics of the difference in the \apml\ rotation curves of young and old stars, and forbidden velocities. We use these to predict the locations where rotation curve reversals can be observed by {\it HST} and the Vera Rubin Telescope. Such measurements would represent support for the amplitude of the bar being a continuous function of age, as predicted by kinematic fractionation, in which the bar strength variations are produced purely by differences in the random motions of stellar populations at bar formation.

\end{abstract}




\section{Introduction}
\label{s:intro}

More than half of the galaxies in the local Universe host a bar \citep{eskridge_etal_00, menendez-delmestre+07, bar_etal_08, aguerri+2009, gadotti09}. Bars play an important role in driving the dynamics and structural properties within the central regions of galaxies via secular processes, including the formation of bulges \citep[see the review by][]{kormendy_13review}. Two bar-driven processes can vertically thicken a bar. The higher radial velocity dispersion due to orbital motion along the bar's major axis makes the bar susceptible to the buckling instability \citep{raha+91, merritt_sellwood94, debattista+06, martinez-valpuesta+06, collier_2020, lokas_2020}. The buckling instability causes the bar to thicken very rapidly. The second mode of vertical thickening is via the trapping of orbits on vertical resonances \citep{combes_sanders81, combes+90, pfenniger_friedli91, quillen02, skokos+02a,  debattista+06, quillen+14}.  This symmetric form of vertical thickening has recently been demonstrated explicitly in $N$-body simulations \citep{sellwood_gerhard_2020}. Unlike the buckling instability, heating by orbit trapping is a slow process.

In both mechanisms, the resulting bulge morphology is boxy or peanut shaped. Such bulges are commonly referred to as boxy/peanut- (B/P) or X-shaped bulges. Stronger features can appear as an X-shape when the bar is viewed edge-on, with its major axis perpendicular to the line of sight (LOS) \citep{athanassoula_misiriotis02,athanassoula05}.  B/P bulges appear in up to 80 per cent of local high mass (\ie\ those with characteristic stellar mass $\log\left(M_{\star}/\mathrm{M}_{\odot}\right) \gtrsim 10.4$) barred galaxies, a fraction that declines rapidly at lower masses \citep{erwin_debattista17}. This characteristic mass appears to have remained unchanged since redshift $z\sim 1$ \citep{kruk+19}.

The \emph{in situ} separation of different populations within a B/P bulge as presented in \cite{debattista+17} demonstrates that co-spatial populations with varying initial radial velocity dispersions evolve separately in a growing bar. As a result, kinematically cooler populations form a strong bar and strongly peanut-shaped bulge, whereas hotter populations form a weaker bar, and are more vertically heated. They termed this process kinematic fractionation. Correlations between kinematics and stellar properties such as age and metallicity during bar formation then result in gradients developing in the final morphology of the B/P bulge and bar \citep[see also][]{fragkoudi+18, debattista+19}. \cite{gonzalez+17} demonstrated that the metallicity distribution of NGC~4710 is more peanut-shaped than the density, as predicted by kinematic fractionation. An alternative mechanism for producing a vertical metallicity gradient relies on the transition between a metal-rich thin disc and a metal-poor thick disc \citep{bekki_tsujimoto11, pdimatteo16}.
This led \cite{pdimatteo+19} to argue that, in addition to the radial velocity dispersions, the vertical dispersion also played a key role in the vertical thickening of populations. However, \citet{debattista+20} showed vertical thickening is a monotonic function of the initial radial action of a given stellar population. Consequently, a thick disc can produce a vertical gradient largely because it has a higher radial velocity dispersion.

The Milky Way (MW) is now understood to host a B/P bulge. Early evidence for this shape was the bimodal density distribution of red clump (RC) stars in the bulge \citep{mcwilliam_zoccali10, saito+11} along the line of sight (LOS) to the Galactic Centre.  This bimodality is produced by the two arms of an X-shape. This structure can be seen directly in the infrared by observing towards the Galactic Centre with {\it Wide-field Infrared Survey Explorer} \citep{ness_lang16}. Various lines of evidence for kinematic fractionation having occurred in the MW have been obtained. \cite{ness+12} demonstrated that the double RC is only traced by metal-rich stars, which was later confirmed with data from \gaia-ESO DR1 and VISTA Variables in Via Lactea \citep[VVV,][]{rojas-arriagada+14} and more recently by \citet{lim+2021} in the Blanco DECam Bulge Survey (BDBS). The behaviour of the RC is the Solar-perspective equivalent of the strongly peanut-shaped metallicity distributions found in external galaxies. \cite{zoccali+17} showed that the 3D density distributions of MW metal-rich and metal-poor stars are boxy and spheroidal, respectively. \citet{catchpole+16} demonstrated an age dependence of the bar strength by considering Mira variables of different periods, showing that the younger Miras trace a stronger bar. \citet{grady+19} also found a similar dependence of bulge morphology on stellar age, with the youngest Miras showing a strong bar with a peanut distribution, which is not seen in the oldest stars. \cite{grady+19} estimated that the bar formed $\sim 8 - 9 \Gyr$ ago, roughly $5 \Gyr$ after the MW formed.

Kinematic studies of the bulge have shown indications of bar streaming motions at low latitudes in both LOS velocities and proper motions \citep{babusiaux+14}. The correlation between the two components, as measured by vertex deviation, indicates the presence of elongated stellar orbits \citep{babusiaux+10, hill+11, vasquez+13}. Measurements of the vertex deviation in Baade's Window show clear non-zero values in metal-rich stars, indicating their stronger bar structure \citep{portail+17,debattista+20}. The dependence of bulge kinematics on chemistry is also seen in the radial velocity dispersion. Metal-rich stars have lower dispersion than metal-poor stars \citep{zoccali+17} except close to the plane ($b \lesssim 1\degrees$), which has been attributed to the central density peak observed by \citet{valenti+16}. The radial velocity dispersion of metal-rich stars decreases steeply away from the centre whereas the gradient in metal-poor stars is much shallower \citep{kunder+12, ness+13b}. For a review of the chemodynamics of the MW bulge, see \citet{bar_chi_ger_18} and references therein. These kinematic differences have been interpreted as further evidence of kinematic fractionation in the bulge \citep{debattista+17, debattista+19}, although the metal-poor stars require a contribution from the stellar halo to explain the observations completely.

\citet[][hereafter C18]{clarkson+18} studied proper motions in the well-observed {\it Sagittarius Window Eclipsing Extrasolar Planet Search} (SWEEPS) field \citep[see also][]{sahu+06, clarkson+08} imaged by the {\it Hubble Space Telescope} ({\it HST}). \citetalias{clarkson+18} used proper motions calculated by \cite{calamida+14} from observations collected over a 10-year baseline with the Advanced Camera for Surveys/Wide Field Camera (ACS/WFC) onboard {\it HST} and derived photometric parallaxes for main-sequence stars. They also used photometry from the Bulge Treasury Survey \citep[BTS,][]{brown+10} to tag photometric metallicities to the stars within the field, allowing them to construct a metal-rich and a metal-poor population of main-sequence stars. They found that the longitudinal proper motion rotation curves (\ie\ \apml\ as a function of distance) were distinct for the two populations. Metal-rich stars exhibited larger amplitude proper motions, with a steeper gradient through the zero point in distance (approximately the Galactic Centre). \citetalias{clarkson+18} suggested that this could be the signature of orbital differences as predicted by kinematic fractionation.

Proper motion rotation curves therefore have the ability to constrain the different kinematic states of bulge populations, and therefore the formation of the bulge. The study of \citetalias{clarkson+18} represents a deep `pencil-beam' along a single LOS close to the bulge's minor axis, but provides little insight into how the proper motion rotation curves vary across the entire bulge. Therefore, in this paper, we explore the trends expected for proper motion rotation curves of different populations in the bulge. We study the proper motions in a star-forming simulation which forms a B/P bulge to predict and interpret trends in the rotation curves across the MW's bulge. The model we use is the same as that in \citet{debattista+17} which they showed had experienced kinematic fractionation.  Therefore, our study will predict the expected trends for proper motion rotation curves if this mechanism has been the main process responsible for the distribution of the MW's stellar populations.

The paper is organized as follows. We describe the model used in this study in Section~\ref{s:simulation}. In Section~\ref{s:separation}, we explore the separation of rotation curves in the SWEEPS field along with a metric we define for quantifying the separation amplitude and present an interpretation of the observed trends with galactocentric velocity maps. In Section~\ref{s:sweepsinterp}, we explore the SWEEPS field in greater detail and compare our results with MW observational data to test the robustness of our separation amplitude measurement. We also provide predictions for the rotation curves in key lines of sight within the MW bulge. In Section~\ref{s:projections}, we explore how galactocentric velocities project onto longitudinal proper motions and define a second measurement of kinematic separation between populations. Section~\ref{s:m2-comp} presents our comparison to a second model with a weaker bar and B/P. Finally, in Section~\ref{s:discussion}, we discuss the implications of our findings and predictions for future work.


\section{Simulation}
\label{s:simulation}

We analyse a high-resolution $N$-body+smoothed particle hydrodynamics (SPH) star-forming simulation which forms a barred spiral galaxy from a hot gas corona embedded in a live dark matter halo. The model has been described numerous times in earlier works where it has been compared to both the MW and external galaxies \citep{cole+14, gardner+14, ness+14, gonzalez+16, gonzalez+17, debattista+17}. \citet{debattista+17} demonstrated that the model underwent kinematic fractionation, and has different bulge (and bar) properties for older (metal-poor) and younger (metal-rich) populations.  The resulting trends are comparable to those seen in the MW.

The initial conditions are comprised of a hot gas corona inside a dark matter halo.
The dark matter halo is comprised of 5 million particles having a force softening of $\epsilon = 103\pc$, virial radius $r_{200}=198 \kpc$ and virial mass $M_{200}=9.0 \times 10^{11} \Msun$. The gas corona consists of 5 million gas particles with a force softening of $\epsilon = 50\pc$. The gas corona has angular momentum $L_z \propto R$ with spin $\lambda \approx 0.041$.

The simulation is evolved using the $N$-body+SPH code {\sc gasoline} \citep{gasoline} with a base time step of 10 Myr. The gas in the corona cools and settles to the centre forming a disc. The formation of a stellar particle happens when gas reaches densities greater than 100 amu $\mathrm{cm}^{-3}$ with a temperature of $T < 15\,000$ K. 10\% of gas in this state forms stars with 35\% the mass of the initial gas particles, corresponding to $\approx9.4 \times 10^3 \Msun$. Gas particles in this state will continue to trigger star formation until their mass falls to below $21\%$ of their initial mass. Then the remaining mass is redistributed to its nearest neighbours, and the gas particle is removed. Each stellar particle is a representation of a \citet{miller_scalo79} initial mass function. %
Feedback from type Ia and type II supernovae is modelled using the blastwave prescription of \cite{stinson+06}. Stellar winds of asymptotic giant branch stars using the theoretical yields for iron and oxygen from \cite{woosley+weaver95} also enrich the interstellar medium. This simulation does not include the diffusion of metals between gas particles \citep{loebman+11} producing the undesirable effect of forming low metallicity stars at all ages, broadening the metallicity distribution.

After $10 \Gyr$ of evolution $\sim 11$ million star particles have formed, with a total mass of $\sim 6.5 \times 10^{10} \Msun$. The resulting disc has a scale length $R_d \approx 1.7\kpc$ \citep{cole+14}. The bar forms between $2 - 4 \Gyr$, after which it continues to grow secularly. We define the bar radius, $r_{bar}$, as the mean of the radii where the amplitude of $m=2$ Fourier moment reaches half its peak value \citep{debattista_sellwood00} and that where the $m=2$ phase angle changes by $10 \degrees$ from constant. At $10 \Gyr$, $r_{bar}\sim 3 \kpc$ \citep{cole+14}.


\subsection{Comparing with the Milky Way}
\label{ss:milkyway}

As shown by \citet{debattista+17}, this model provides insights into trends in the MW, and makes predictions which can be tested against current and future observations. By scaling the $t = 10 \Gyr$ time-step as in \citet{debattista+17}, we can produce a bar of about the right size with roughly the correct kinematics. Here we describe how we scale the model and qualitatively compare to the MW.

We spatially scale up the simulation by a factor of 1.7, in line with recent measurements for the MW's bar length, $r_{bar} = 5.0~\pm~0.2 \kpc$ \citep{wegg+15}. After rescaling, the model's bar length is $r_{bar} = 4.85~\pm~0.55 \kpc$. The velocities are scaled by 0.48 to match the velocity dispersion in the MW bulge \citep[see][]{debattista+17}. We place the observer at $8\kpc$ from the galactic centre in the mid-plane with the bar aligned to $27 \degrees$ from the LOS of the observer to the galactic centre \citep{wegg_gerhard13, qin+18}, with the near side of the bar at positive longitude.

In order to increase our resolution in the bulge region, we assume the simulation to have mid-plane symmetry; therefore, we project stars below the plane onto above the plane with an inverted vertical velocity ($z' = -z$; $v_z'=-v_z$, for $z<0$). We then calculate galactic longitude, latitude, and LOS distance $(l,b,D)$ along with longitudinal proper motions $(\pml)$ for each star from the solar perspective, in the Galactic rest frame. Coordinate transformations in this work were computed using the {\sc PYTHON} package {\sc GALPY} \citep{galpy}.

We define the bulge region as follows: $|l|\leq20\degrees$, $2\degrees\leq|b|\leq10\degrees$, and $5.75 \leq D/\kpc \leq 10.25$. This is larger than previous studies which usually constrain longitude to $|l|<10\degrees$. Considering that the bar is inclined by $27\degrees$, at $l=+20\degrees$, we sample $\sim 3.5\kpc$ along the near side of the bar encapsulating a larger extent of the B/P component. Our range is also larger than the proposed footprint of the Vera Rubin Observatory, Legacy Survey of Space and Time (LSST) bulge observations allowing for predictions for this and additional future survey missions (see Section~\ref{ss:futurepros}).

\begin{figure}
	\centerline{
		\includegraphics[angle=0.,width=\hsize]{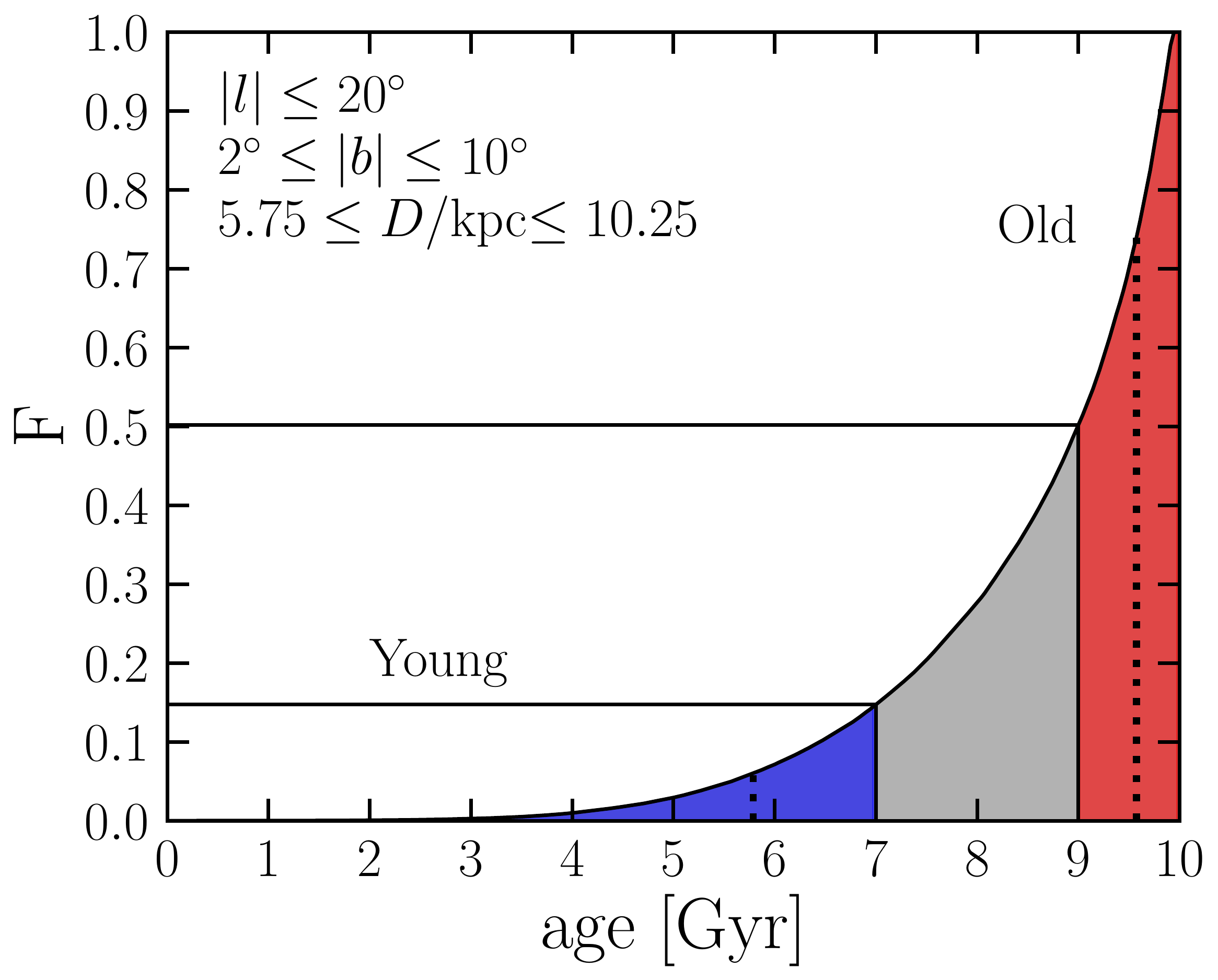}\
	}
	\caption{The cumulative distribution of ages within the model's bulge region at $t = 10 \Gyr$. The spatial cuts used are given at the top left-hand side. We define the young (blue) and old populations' (red) age cuts as $7$ and $9\Gyr$, respectively. The mean ages for the two populations (vertical black dashed lines) are $5.8=$ and $9.6\Gyr$, respectively.}
	\label{fig:agedist}
\end{figure}

The lack of chemical mixing in this model results in an excess of stars with low metallicities at all ages. \citet{debattista+17} circumvented this problem by considering stellar populations defined by age, rather than by metallicity. Likewise, here we also define populations based on stellar ages. The cumulative distribution of ages within the model's bulge is shown in Fig.~\ref{fig:agedist}.  Using this distribution, we define a young population as those stars with age $<7 \Gyr$ (mean age = 5.8\Gyr) and an old population with stellar ages $>9 \Gyr$ (mean age = 9.6\Gyr). This results in a sample of $361\:131$ relatively young stars and $1\:341\:922$ old stars within our defined bulge region, representing 14 and 51 per cent of all bulge stars, respectively. Since the simulation was run for only $10\Gyr$ the majority of ages are lower than would be expected for the MW; none the less, the ordering of the stellar ages would remain intact. These age ranges allow us to qualitatively compare the simulation with populations separated by metallicities in the SWEEPS field \citep{bernard+18} and the MW bulge in general. We also note that the distribution of ages in this model is consistent with the picture of a largely old bulge in the MW \citep{kuijken_rich02, zoccali+03, clarkson+08, clarkson+11, brown+10, valenti+13, renzini+18}, as discussed in \citet{debattista+17}. The model also has a similar fraction of stars younger than $5 \Gyr$ observed in the MW \citep[$\sim 3 \%$ of the bulge population][]{renzini+18}. While we refer to a young population in the model's bulge, we mean this in a relative sense: even excluding that the model is only evolved for $10 \Gyr$, a large majority of the young stars are old with $\sim 50\%$ of them formed during the bar's formation.

\begin{figure}
	\centerline{
		\includegraphics[angle=0.,width=\hsize]{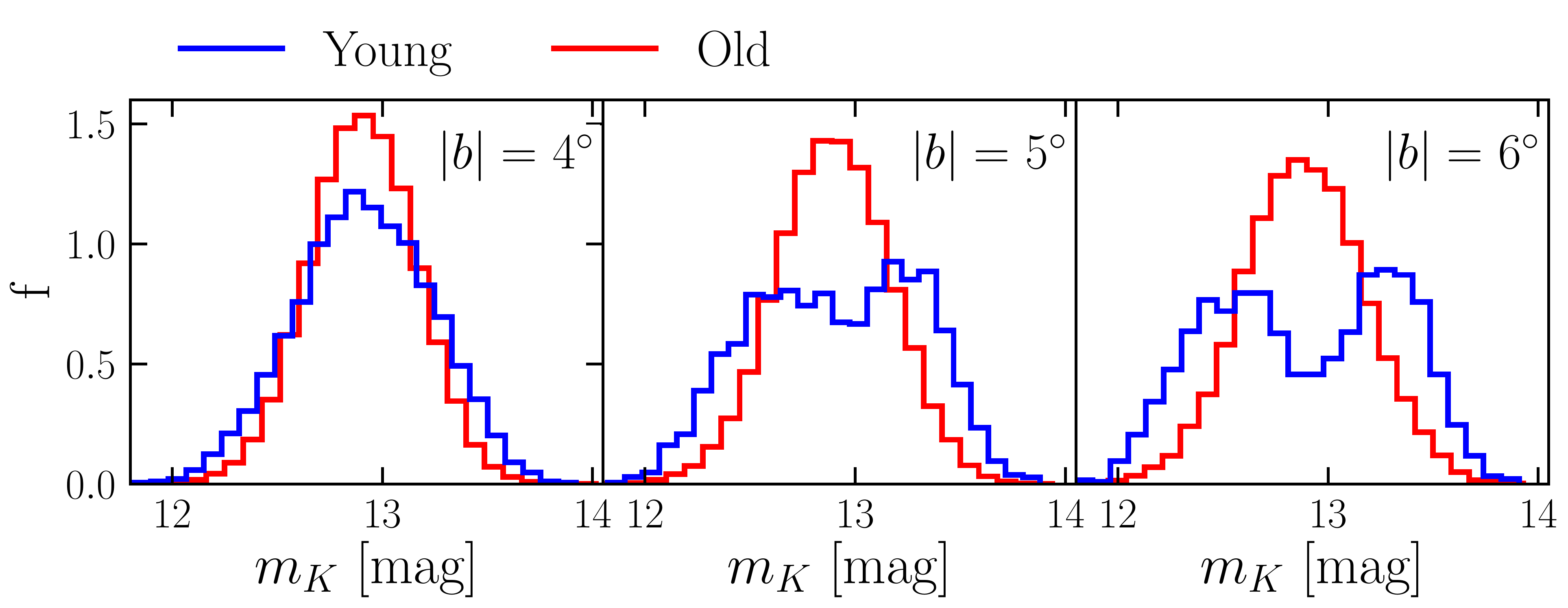}\
	}
	\caption{Unextincted apparent magnitude distributions of simulated RC stars along the LOS within $|l|<4\degrees$ for $|b|=4\degrees$ (left-hand panel), $5\degrees$ (middle panel) and $6\degrees$ (right-hand panel) with $\delta |b| = 0.25\degrees$. Young (age $<7\Gyr$) and old (age $>9\Gyr$) stars are represented by the blue and red histograms, respectively.  The magnitude distributions have been convolved with a Gaussian of width $\sigma=0.17~{\rm mag}$ to represent the width of the RC. As in the MW, a bimodality is first evident at $|b| \simeq 5\degrees$.}
	\label{fig:doublerc}
\end{figure}

We verify that the vertical structure of our rescaled model is a reasonable analogue of the MW's B/P bulge by considering the variation of the distance bimodality as a function of latitude, as viewed from the Sun.  In particular, we consider the double RC as a function of latitude. Following the similar prescription of \citet{gonzalez+15} and \citet{debattista+17}, we assume that the RC stars follow the same density as the model in general. We therefore set the absolute magnitude of all stars to the average of the RC, $M_K=-1.61$, and convert this to apparent magnitudes, $m_K$, based on their distance from the solar position ($8 \kpc$). We then convolve each $m_K$ with a Gaussian kernel of $\sigma=0.17~{\rm mag}$ to approximate the width of the RC magnitude distribution \citep{gerhard_martinez-valpuesta12}. We present the magnitude distribution of simulated RC stars split by our age cuts in Fig.~\ref{fig:doublerc}. The distribution of young stars is single peaked at $|b|=4\degrees$ and bimodal above that, in agreement with the bimodality found by \citet{ness+12}. The old population is single peaked at all latitudes.

\section{Separation of rotation curves}
\label{s:separation}

\begin{figure*}
	\centerline{
		\includegraphics[angle=0.,width=\hsize]{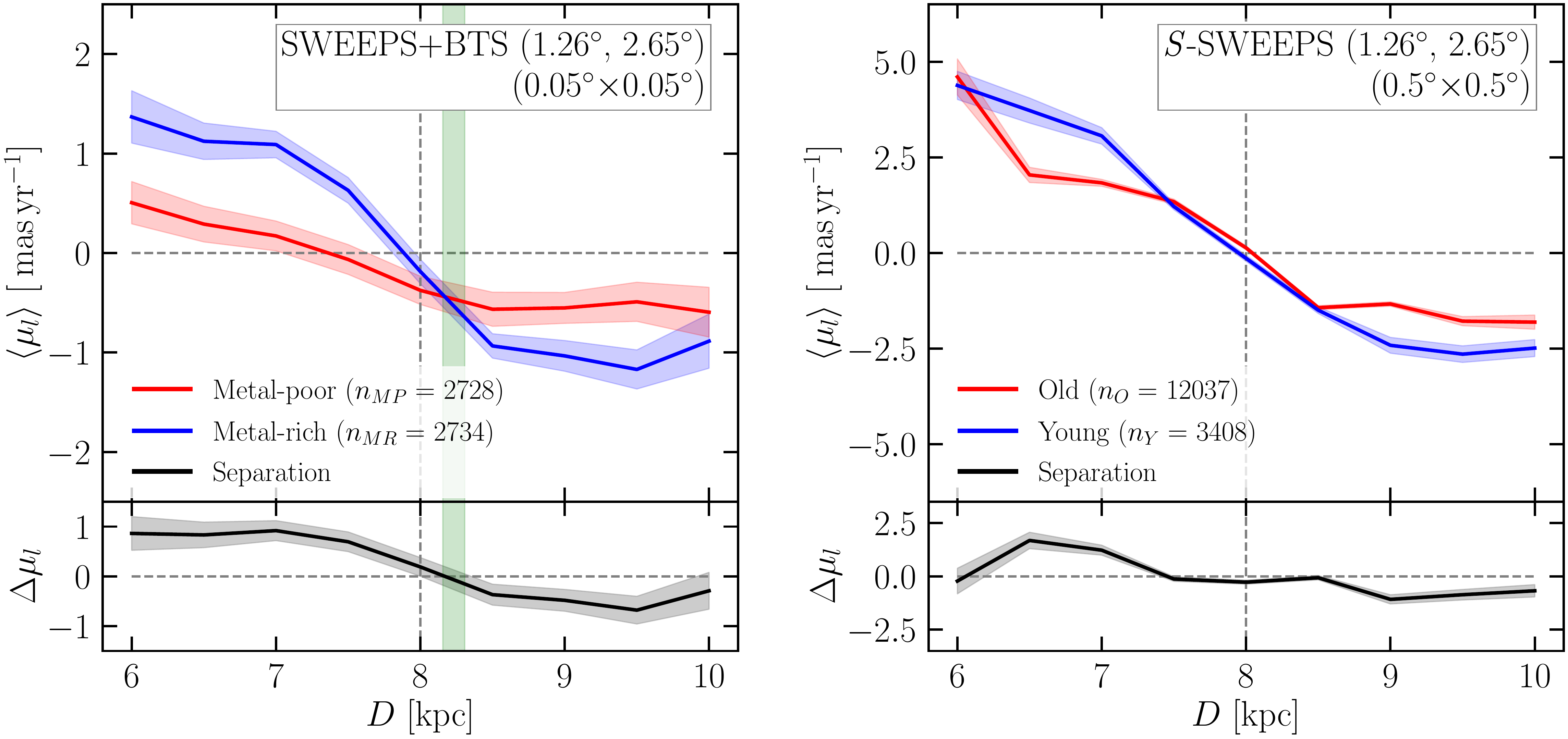}\
	}
	\caption{Left-hand panel: Average longitudinal proper motion rotation curves for metal-rich and metal-poor main-sequence stars of the SWEEPS+BTS field, centred on $(l,|b|)=(+1.26\degrees{},2.65\degrees{})$ with an FOV of $0.05 \times 0.05 \, {\rm deg}^2$ (\citetalias{clarkson+18}). The vertical green shaded region show the range of estimates of $R_\odot$ from the GRAVITY consortium. Right-hand panel: Average longitudinal proper motion rotation curves for young and old stars, and the separation between them, of the simulated {\it S}-SWEEPS field, centred on $(l,|b|)=(+1.26\degrees{},2.65\degrees{})$ with an FOV of $0.5\degrees \times 0.5\degrees$.  The number of star particles in each population is listed in the legends.
		\label{fig:LOS-SWEEPS}}
\end{figure*}

The combined field from SWEEPS+BTS data studied in \citetalias{clarkson+18} was approximately $3.4 \times 3.4 \, {\rm arcmin}^2$ centred at $(l,b)_{J2000.0}\approx(+1.26\degrees,-2.65\degrees)$. We compare the rotation curves of young and old stars in the model's equivalent of the SWEEPS+BTS field, hereafter {\it S}-SWEEPS\footnote{We use the prefix ``{\it S}-'' throughout this paper to denote simulated equivalents of observed {\it HST} fields.}. Although the model has a large number of star particles, the overall number is still small compared to the real MW; therefore our field of view (FOV) is increased to $30 \times 30 \, {\rm arcmin}^2$ to increase the number of particles. As mentioned in Section~\ref{ss:milkyway}, we constrain our distance measurements within the bulge to between $5.75$ to $10.25 \kpc$. Taking bins along the LOS, we calculate the mean longitudinal proper motion for young and old stars, $\apml_{Y}$ and $\apml_{O}$, within each bin, with the standard error given by
\begin{equation}
   e_{\apml} = \frac{\sigma_{\mu_l}}{\sqrt{n_{\star}}}.
   \label{e:apmlerror}
\end{equation}
The rotation curve separation between the young and old populations is then simply $\Delta\mu_l = \apml_{Y}-\apml_{O}$ in each bin and the uncertainty, $e_{\Delta\mu_l}$, is propagated through addition in quadrature.

We explore the SWEEPS+BTS data using the same binning as described above. The left-hand panel of Fig.~\ref{fig:LOS-SWEEPS} presents the sample from \citetalias{clarkson+18} (their figure 8), showing the average longitudinal proper motion as a function of LOS distance for their metal-rich versus metal-poor main-sequence populations. The distances are estimated from photometric parallax, using as reference the median distance modulus $(m-M)_0=14.45$ of the SWEEPS+BTS field main sequence \citep{calamida+14}, which, taken literally, corresponds to a physical distance of $D_0 = 7.76 \kpc$. This is the distance of the median well-measured population that survives their kinematic cut for bulge objects ($\mu_l < -2$~mas yr$^{-1}$), and thus naturally lies closer to the observer than the Galactic Centre. The offset between the proper motion zeropoints $\mu_l=0$~in the two panels arises because the SWEEPS+BTS proper motions are measured relative to the median well-measured (majority-bulge) stellar population in the FOV (with median distance closer than $8\kpc$; see \citetalias{clarkson+18} and \citealt{calamida+14} for more on this issue), whereas for our simulated samples, $\mu_l=0$~at the galactic centre by construction. This offset in proper motion zeropoint does not impact our results in any way.

The identification of photometric parallax with physical distance gains some support when we consider the distance at which the `metal-rich' and `metal-poor' rotation curves cross, which is approximately the same as the recent measurement of the Galactic Centre distance by the GRAVITY experiment, at $R_\odot \approx 8.156 - 8.308 \kpc$~\citep{GRAVITY+19,GRAVITY+21}, shown as the green shaded region in the left-hand panel of Fig.~\ref{fig:LOS-SWEEPS}.

The rotation curves from the \textit{S}-SWEEPS field are shown in the right-hand panel of Fig.~\ref{fig:LOS-SWEEPS}. For both the simulation and observations we show the rotation curve separation, $\Delta\mu_l$ below each panel. In both the young and old populations, $|\apml|$ rises on either side of the galactic centre. The peak value of $|\apml|$ on the near side is larger than that on the far side, by a factor of about 2 for both populations, which is expected because of perspective. The ratio of peak amplitude of young stars to old stars is also $\sim 2$. The largest $\Delta\mu_l$ is at $\sim1\kpc$ from the galactic centre. These results are qualitatively similar to those of \citetalias{clarkson+18} for the metal-rich vs. metal-poor main-sequence populations shown in the left-hand panel.


\subsection{Separation amplitude}
\label{ss:sepamp}

Given that the model matches the trends found by \citetalias{clarkson+18}, we consider the behaviour of the rotation curves of young and old stars across the model's bulge, to predict trends that can be tested in future studies.

\begin{figure*}
	\centerline{
		\includegraphics[angle=0.,width=.95\hsize]{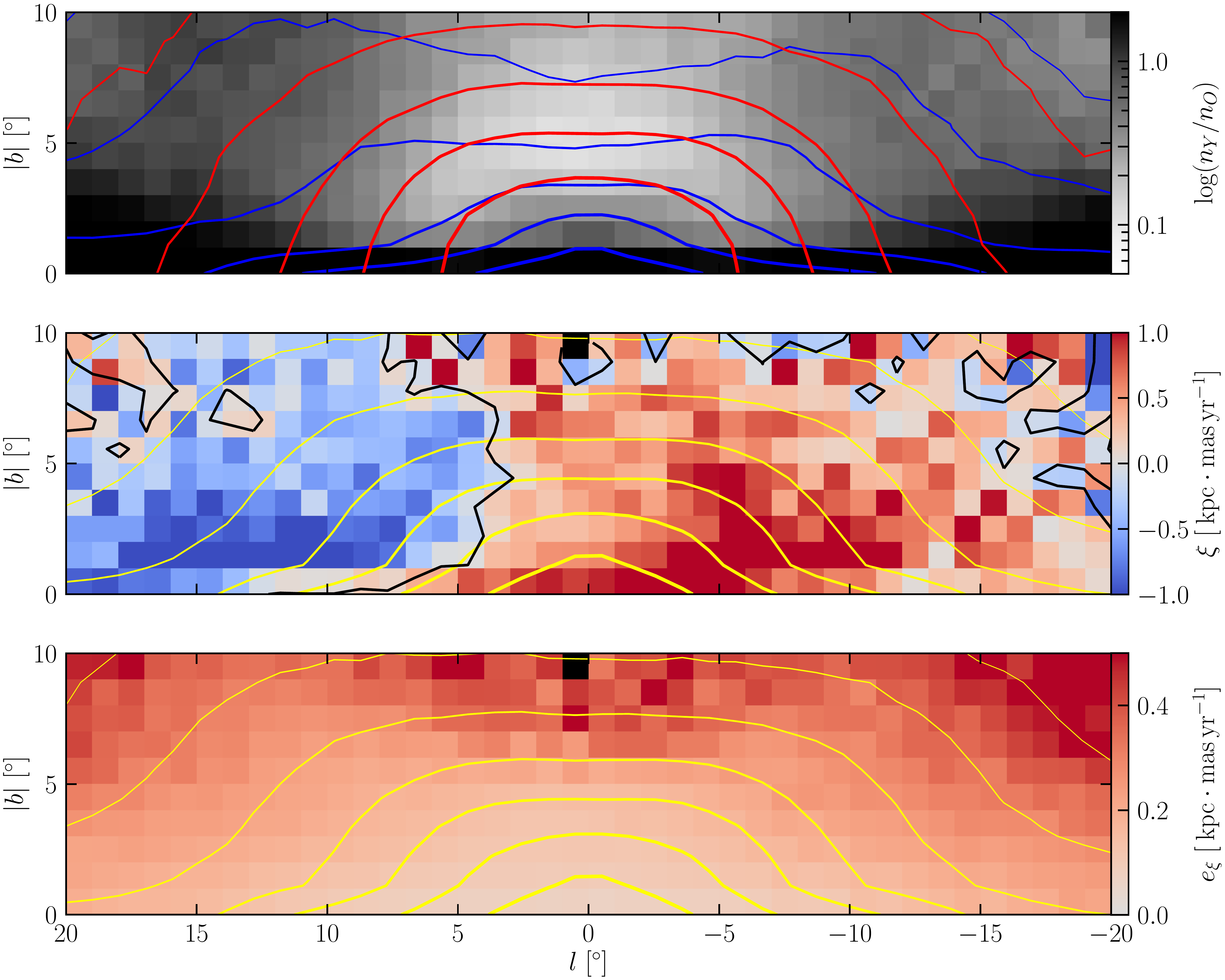}\
	}
	\caption{Top panel: fractional distribution of young to old stars within the bulge of the model. Blue and red contours follow young and old population densities, respectively. Middle panel: separation amplitude, $\xi$, for each pixel representing a $1 \times 1 \, {\rm deg}^2$ field. Bottom panel: uncertainty on $\xi$ for each field. In the bottom two panels, the yellow contours follow the density of all bulge stars. Black pixels are fields for which $\xi$ could not be measured reliably.
		\label{fig:field-sepamp}}
\end{figure*}

The top panel of Fig.~\ref{fig:field-sepamp} shows, in Galactic coordinates $(l,|b|)$, the density distribution of stars in the two populations. The distribution of young population (blue contours) is more pinched at high latitude, resembling a peanut, whereas the old population (red contours) appears more boxy. The young stars are also more concentrated to the mid-plane, demonstrating there are fewer young stars at higher latitudes, similar to the metallicity distributions \citep{zoccali+17} and the distribution of long-period (young) Miras \citep{grady+19} within the MW.

To quantify the separation between the young and old rotation curves, we integrate the separation along the LOS, to obtain a separation amplitude. Binned by distance, we define the separation amplitude $\xi$,  as the LOS integral of $\Delta\mu_l(D)=\apml_{Y}(D)-\apml_{O}(D)$
as
\begin{equation}
   \xi = \delta D \cdot \sum_{D=d_1}^{d_2}\Delta\mu_l(D),
   \label{eq:xi}
\end{equation}
where $D$ is each distance bin centre with width $\delta D = 0.5\kpc$ and $\xi$ has units of \kpcmasyr.
We set $d_1=6\kpc$, $d_2=10\kpc$, respectively, as the limits of the model's bulge region.

We map $\xi$ across the entire bulge of the model, using fields of $1 \times 1 {\rm deg}^2$. This represents a much larger FOV than those sampled by deep bulge fields in the MW, but is necessary to attain reasonable particle numbers at higher latitudes. For each $(l,|b|)$ bin, we repeat the analysis applied to the {\it S}-SWEEPS field, producing \apml\ rotation curves for the young and old populations using the same binning along each LOS.

The middle panel of Fig.~\ref{fig:field-sepamp} shows a map of $\xi(l,|b|)$ for the model. We focus on the region $|b| > 2\degrees$ to avoid the thin disc and the nuclear disc found below this latitude in our model \citep{cole+14, debattistaMWND+15, debattistaMWND+18}.
Along the minor axis, $|l| \lesssim 5\degrees$, $\xi$ is mostly positive up to large latitudes with relatively low amplitudes, $0~<~\xi/\kpcmasyr~<~0.5$. These rotation curves have qualitatively similar separation profile to the {\it S}-SWEEPS field (Fig.~\ref{fig:LOS-SWEEPS}) which is more or less antisymmetric with distance from the galactic centre, resulting in relatively small $\xi$ values. The small positive values of $\xi$ on the minor axis arise largely because of perspective.

As we show below, some rotation curves of fields away from the minor axis are not anti-symmetric, resulting in separation profiles that are everywhere positive or negative; for these rotation curves, $\xi$ will be larger. Almost all fields with longitudes $l<5 \degrees$ have positive $\xi$ values, whilst for longitudes $l > 5\degrees$ $\xi$ has mostly negative values. Away from the minor axis, there is a slight vertical gradient with higher amplitude $\xi$ values at low latitude which decreases with increasing latitude.

The number of young stars decreases rapidly with increasing height, and above $|b| \gtrsim 8\degrees$ some fields have too few young stars to measure a reliable rotation curve. We calculate the uncertainty of the separation amplitudes, $e_\xi$, for each field as
\begin{equation}
   e_{{\xi}} = \delta D \cdot \left( \sum_{D=d_1}^{d_2}  {e_{\Delta\mu_l}(D)}^2 \right)^{1/2},
   \label{e:sepamperror}
\end{equation}
where $\delta D = 0.5\kpc$ is the bin width.

The distribution of $e_\xi$, presented in the bottom panel of Fig.~\ref{fig:field-sepamp}, loosely traces the density distribution of young stars (blue contours in the top panel), highlighting that the number of star particles along an LOS is a limiting factor in this measurement. The uncertainties are lowest on the minor axis and on the near side of the bar.


\subsection{Galactocentric velocities}
\label{ss:bulgekin}

\begin{figure*}
	\centerline{
		\includegraphics[angle=0.,width=\hsize]{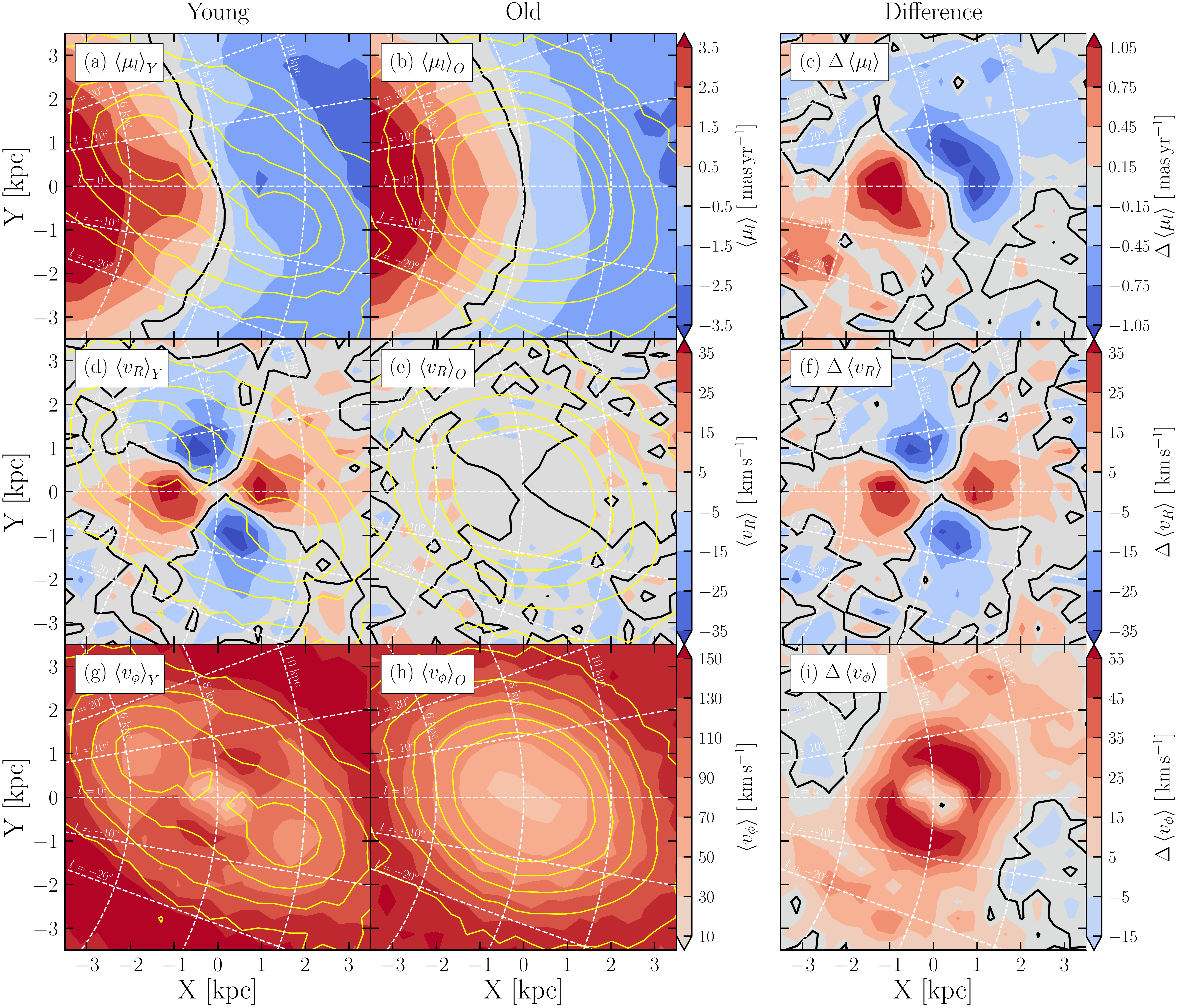}\
	}
	\caption{Average velocity fields in the $(X,Y)$ plane for stellar particles at $0.5<|z|/\kpc<1.0$. The left-hand column presents the kinematics of the young population while the middle column shows those of the old. The difference in the velocity fields between young and old stars is shown in the right-hand column.  The top row shows heliocentric longitudinal proper motions, the middle row shows galactocentric radial velocities and the bottom row shows galactocentric tangential velocities. Yellow contours follow log densities of the corresponding population. Black contours indicate where each velocity component equals zero. White circular dashed lines outline distances $6$, $8$ and $10 \kpc$, while the white straight dashed lines mark longitudes between $20\degrees$ and $-20\degrees$ in $10\degrees$ intervals. The observer is at $X = -8\kpc$ in this figure.
		\label{fig:XY-grid-sep}}
\end{figure*}

In order to interpret the $\xi$ map, including the asymmetries between positive and negative longitudes, we consider the difference in the bulge's intrinsic (galactocentric cylindrical) kinematics, \ie\ the galactocentric radial velocity, $v_R$ and galactocentric tangential velocity, $v_{\phi}$. Fig.~\ref{fig:XY-grid-sep} first presents the vertically averaged heliocentric longitudinal proper motions for the bulge's young and old populations in the $(X,Y)$ plane, along with the corresponding galactocentric cylindrical velocities. We only consider stars in the vertical slice $0.5<|z|/\kpc<1.0$, equivalent to $4\degrees < |b| < 7\degrees$ at $8 \kpc$, to avoid the effect of the nuclear disc as discussed in Section~\ref{ss:sepamp} and regions where the uncertainty in $\xi$ is largest. The left-hand and middle columns of Fig.~\ref{fig:XY-grid-sep} show the velocity distributions of the young and old populations respectively. The right-hand column shows the difference between the young and old velocity distributions. Within each panel, a thick black contour traces the zero amplitude line of each velocity component and where the difference between populations is also zero.
The density of the two populations, indicated by the yellow contours in the left-hand and middle panels, shows that the young stars trace a strongly barred morphology, which has two peaks on either side of the galactic centre. The peaks are the lower layer of the X-shape in the B/P bulge, as also seen in the MW \citep[][their figure 19]{sanders+19}. The old population, instead, is considerably less elongated, supporting only a weak bar, as shown by \citet{debattista+17}.

Panels (a) and (b) show that as a result of the much stronger bar in the young population, their average heliocentric proper motions ($\apml_Y$) exhibit a stronger longitudinal variation relative to the Sun than those of the old stars. The $\apml_Y$ distribution has two high-amplitude regions along the $l=0\degrees$ direction, $1\kpc$ in front of, and $1\kpc$ behind the galactic centre. The black contour in the $\apml_Y$ profile is twisted towards the bar major axis, away from the $d_\odot = 8 \kpc$ line, unlike the $\apml_O = 0$ contour which traces the $8 \kpc$ line more closely at central longitudes. Fig.~\ref{fig:XY-grid-sep}(c) shows the difference in the distribution of \apml\ between the two populations. This panel shows that most of the signal in $\Delta\apml$ comes from regions where the amplitude of $\apml_Y$ peaks, close to $l=0\degrees$. The near peak has a tail towards negative longitude whereas the far peak has a tail to positive longitude. A field close to the minor axis will intersect both the near and far peaks of $\Delta\apml$ which have positive and negative values respectively, resulting in $\xi \sim 0$. From the $(X,Y)$ perspective, we can see that lines of sight away from the minor axis only intersect one of these $\Delta\apml$ peaks, and therefore have larger $|\xi|$ values.

We then turn to the intrinsic kinematics in galactocentric cylindrical coordinates, which is the natural frame of the bar, removing the effects of perspective. In the middle row of Fig.~\ref{fig:XY-grid-sep} we present the distributions of galactocentric radial velocity, $\avg{v_R}$. The young population (Panel d) exhibits a quadrupole pattern, with zero velocity lines (black contours) aligned with the bar major and minor axes. The amplitude of $\avg{v_R}$ peaks at $\sim \pm 45\degrees$ relative to the bar indicating bar-aligned motions along either side of the galactic centre. The old population has no quadrupole pattern and near-zero $\avg{v_R}$ values, reflecting its weaker bar morphology. The resulting difference map, $\Delta\avg{v_{R}}$ (Panel f), therefore also has a strong quadrupole pattern.

The bottom row of Fig.~\ref{fig:XY-grid-sep} presents the distribution of the galactocentric tangential velocity, $\avg{v_{\phi}}$. For the old population (Panel h), the distribution of $\avg{v_{\phi}}_O$ is mildly elongated along the bar. In contrast, the $\avg{v_{\phi}}_Y$ distribution (Panel g), is elongated more strongly whilst also exhibiting a complex inner structure. The lower levels of the X-shape, identified by the density contours, coincide with regions of low $\avg{v_{\phi}}$ positioned approximately $3\kpc$ along the bar major axis. The lowest values of $\avg{v_{\phi}}_Y$ are at the very centre.  Peak values of $\avg{v_{\phi}}_Y$ are along the bar's minor axis, as expected for streaming motions in a highly elongated population. Consequently, the difference between the two populations, $\Delta\avg{v_{\phi}}$ (Panel i), exhibits two peaks along the minor axis.  Regions of $\Delta\avg{v_{\phi}} < 0$ appear slightly beyond $2\kpc$ along the bar major axis, where the young stars are reaching the apocenter of their elongated orbits and the old population has comparable or larger velocities.

Taken as a whole, these differences in the intrinsic kinematics are as expected for the two populations, one strongly tracing the bar (the young one) and one that traces it weakly (the old one). From the Solar perspective, the proper motion rotation curves are then a position-dependent combination of these two motions. The differences between the velocity distributions can be understood largely in terms of the different bar strengths, which themselves are a result of the different random motions of stars at the time of the bar's formation \citep{debattista+17}.


\section{Interpretation of the SWEEPS Field}
\label{s:sweepsinterp}

We now explore the simulated {\it S}-SWEEPS field (Fig.~\ref{fig:LOS-SWEEPS}) in greater detail. Using Eqn.~\ref{eq:xi} we calculate the separation amplitude of the {\it S}-SWEEPS field to be $\xi~=~0.05\pm0.48 \>{\rm kpc \cdot \rm mas\, yr^{-1}}$, commensurate with the regions surrounding this LOS in Fig.~\ref{fig:field-sepamp}. We use the insight derived from Fig.~\ref{fig:XY-grid-sep} to interpret this value in terms of the intrinsic velocities and the resulting rotation curves. Furthermore, we directly compare our model with MW data to test the level of confidence of our model as an approximation of the MW.

\begin{figure}
	\centerline{
		\includegraphics[angle=0.,width=\hsize]{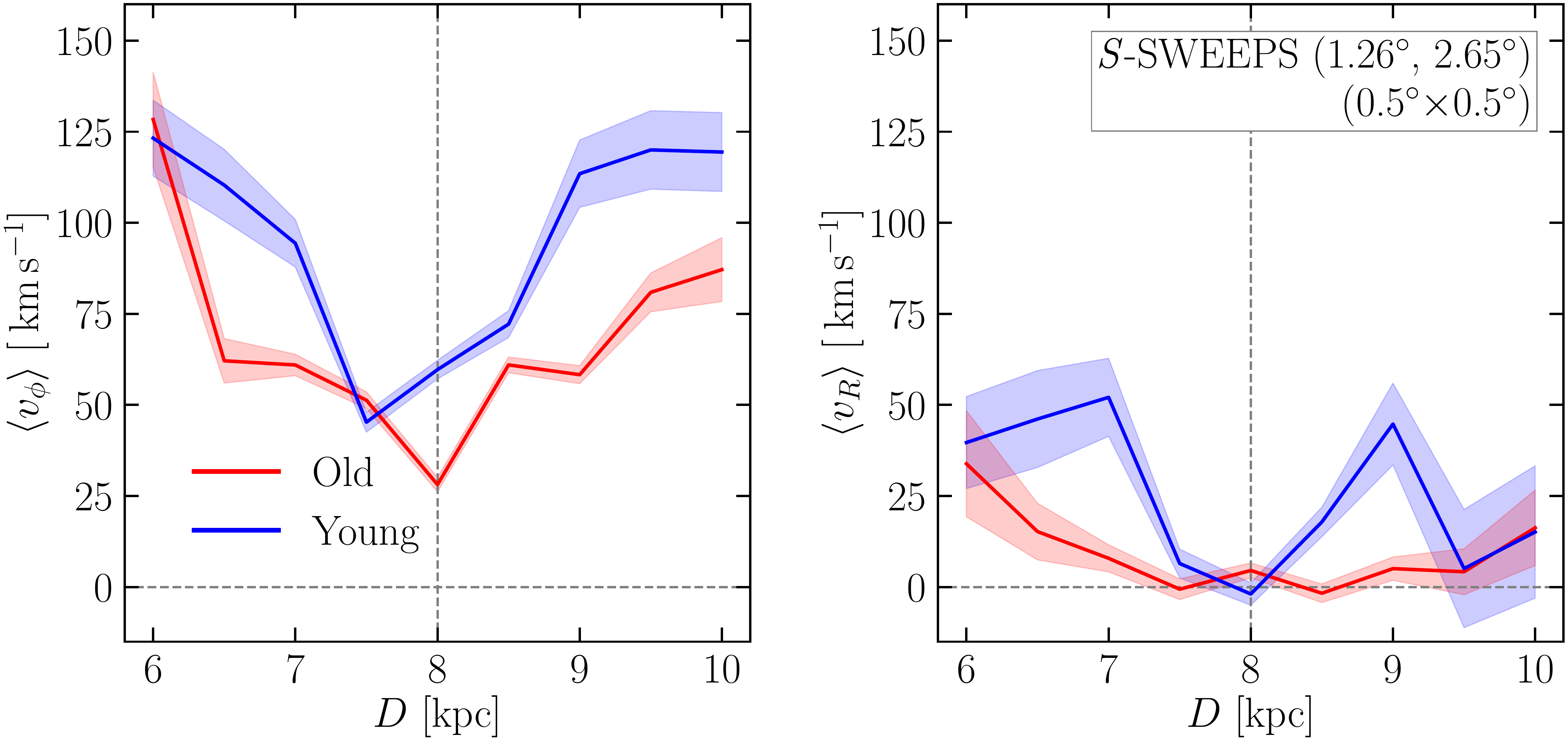}\
	}
	\caption{LOS profiles of galactocentric $\avg{v_{\phi}}$ (left-hand panel) and $\avg{v_{R}}$ (right-hand panel) for the {\it S}-SWEEPS field in the model.
		\label{fig:LOS-SWEEPS-comps}}
\end{figure}

For the {\it S}-SWEEPS LOS, Fig.~\ref{fig:LOS-SWEEPS-comps} shows the distance profiles of galactocentric tangential $\avg{v_{\phi}}$ and galactocentric radial $\avg{v_{R}}$ velocities using the same distance bins as the rotation curves.
Both populations have their lowest $\avg{v_{\phi}}$ close to $8 \kpc$, increasing away from the centre. At almost every distance bin, the young population has the higher  $\avg{v_{\phi}}$. In Fig.~\ref{fig:XY-grid-sep}, we showed that the young population has a stronger bar and kinematics consistent with elongated bar orbits; the peaks of $\avg{v_{R}}$ seen in the right-hand panel of Fig.~\ref{fig:LOS-SWEEPS-comps} support this, and indicate stars moving away from the galactic centre. The old population has low values of $\avg{v_{R}}$ along the LOS, which reflects their more axisymmetric distribution. These profiles show that the differences in rotation curves between the young and old populations are a consequence of differences in both $\avg{v_{\phi}}$ and $\avg{v_{R}}$ because only the young population is strongly barred.


\subsection{Monte Carlo simulation of MW data}
\label{ss:mw-mc}

To compare the separation amplitude of the {\it S}-SWEEPS field with observational data, we apply our methodology to the SWEEPS+BTS data presented in \citetalias{clarkson+18}. We note here that the populations within the model represent the ends of the age distribution (see the coloured regions in Fig.~\ref{fig:agedist}), whereas \citetalias{clarkson+18} split the photometric metallicity distribution within their data using auto-GMM clustering. The estimated mean metallicities of the `metal-rich' and `metal-poor' samples are $\feh_0 \approx -0.24$ and $\feh_0 \approx +0.18$, respectively (see section 3.5 of \citetalias{clarkson+18}). As a consequence of their methodology and the nature of separating by metallicity, the age distributions of their sub-samples may partially overlap.

We use the \citetalias{clarkson+18} data, which the authors used to produce their Fig. 8, showing that metal-rich stars have higher amplitude \apml\ rotation curves than metal-poor stars. For their metal-rich and metal-poor populations, we bin the data in distance following the prescriptions in Section~\ref{s:separation} and calculate the separation amplitude using the same method as we used for the model.

\begin{figure}
	\centerline{
		\includegraphics[angle=0.,width=\hsize]{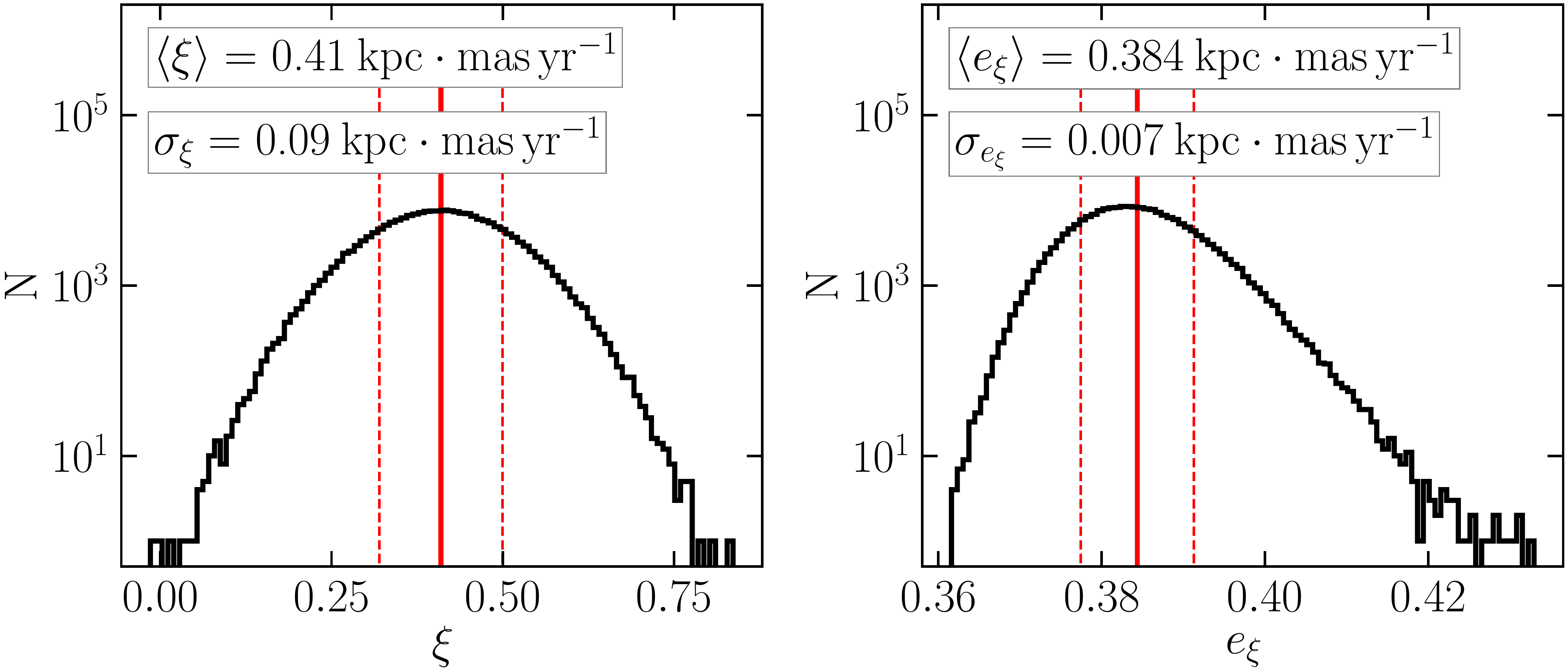}}
	\caption{Results of the Monte Carlo runs to calculate the separation amplitude for the SWEEPS+BTS data from \citetalias{clarkson+18}. Thick red lines show the mean $\xi$ and $e_{\xi}$ values. Red dashed lines denote $1 \sigma$ deviation. At this field, the fiducial model predicts $\xi = 0.05\pm0.48 \>{\rm kpc \cdot \rm mas\, yr^{-1}}$.
		\label{fig:hst-mc}}
\end{figure}

To determine the effect of observational uncertainties on the calculated values and the robustness of the uncertainty estimates ($e_{\xi}$), we run a Monte Carlo (MC) simulation of our separation amplitude measurement. We assume that the uncertainty of the observed longitudinal proper motion to be $\sigma_{\mu_l} = 0.08 \masyr$ (Fig. 17 of \citetalias{clarkson+18}). The photometric parallax used in the distance determination has an estimated uncertainty of $0.119$ and $0.153$~mag for the metal-rich and metal-poor sample, respectively (Table 11 of \citetalias{clarkson+18}). For each run of the MC, we add random errors to the magnitudes and proper motions of each star in the SWEEPS+BTS sample, assuming the error distributions are Gaussian. We then recalculate $\xi$ and $e_{\xi}$ for each run.
Our MC of 200~000 runs produces the distributions of $\xi$ and $e_{\xi}$ shown in Fig.~\ref{fig:hst-mc}. The mean separation amplitudes and errors from the MC runs are $\avg{\xi} = 0.41 \>{\rm kpc \cdot \rm mas\, yr^{-1}}$ and $\avg{e_{\xi}} = 0.384 \>{\rm kpc \cdot \rm mas\, yr^{-1}}$. The value of $\avg{e_{\xi}}$ is relatively large as in the model, however it is well constrained with a standard deviation of $\sigma_{e_{\xi}} = 0.007 \>{\rm kpc \cdot \rm mas\, yr^{-1}}$. The definition of $e_{\xi}$ from Eqn.~\ref{e:sepamperror} corresponds to the sum in quadrature of the uncertainties for each distance bin, where the uncertainty in each bin is given by the proper motion dispersion and the number of stars in the bin. Since the number of stars within the whole distance range ($5.75-10.25 \kpc$) does not change substantially between MC iterations, and the $\mu_l$ uncertainty is low overall, this leads to $\avg{e_{\xi}}$ being well constrained, with a small standard deviation.

Comparing the $\xi$ values of the observational data to our model, we can see that they agree within the uncertainty. Although this is a comparison for a single field on the minor axis, it demonstrates that our metrics can applied to observational data and the model provides a reasonable basis for comparing to the MW. For the SWEEPS+BTS field, the profile is antisymmetric resulting in a low value of $\xi$. Our model has a value of $\xi$ closer to zero and a larger associated error than the MW data. We stress, however, that there are many differences between the observational and simulated measurements so we limit our comparisons to qualitative trends.


\subsection{Other fields}
\label{ss:simfields}

\begin{figure*}
	\centerline{
		\includegraphics[angle=0.,width=\hsize]{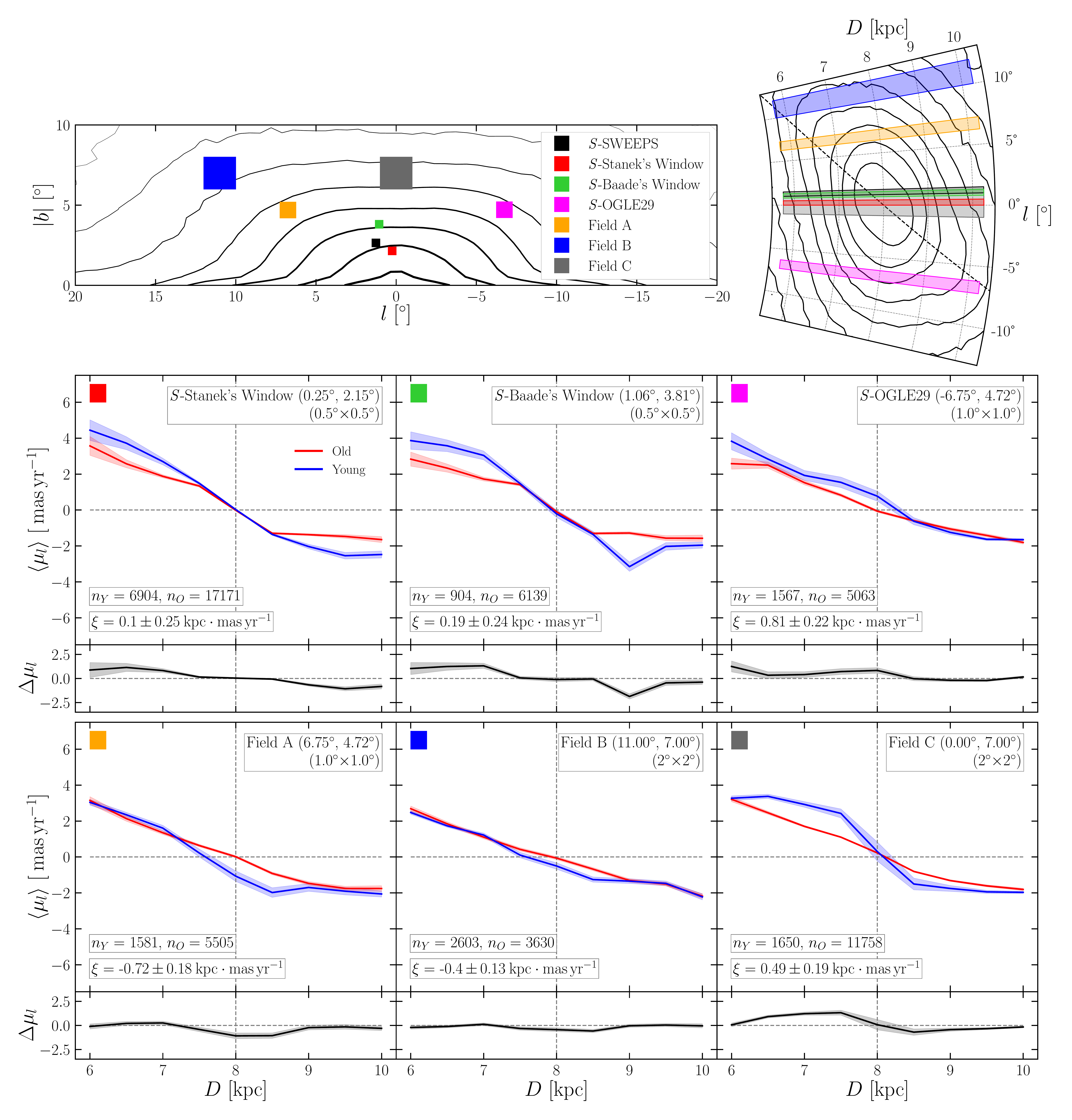}}
	\caption{Top left-hand panel: the positions of the six fields of interest within the model's bulge from the heliocentric perspective. The coloured squares correspond to the sizes used in our simulated fields to capture enough star particles. Black contours follow the log density of all bulge stars. Top right-hand panel: the six fields of interest presented in a top-down view of the model's bulge. The bar major axis is indicated by the dashed line. Black contours follow the log density of all bulge stars. Bottom panel: average longitudinal proper motion rotation curves and the separation for the fields of interest. The field names and FOV are labelled at the top right-hand side. The number of star particles in both populations is also listed along with the calculated separation amplitude, $\xi$.
		\label{fig:additional-fields}}
\end{figure*}

We now expand our analysis to fields for which data are available from {\it HST}-BTS observations. The three remaining BTS fields that can be used for a study similar to that of \citetalias{clarkson+18} are {\it Stanek's Window}, {\it Baade's Window} and the OGLE29 field \citep{brown+10, renzini+18}. We explore comparable fields within our model, which we refer to as {\it S-Stanek's Window}, {\it S-Baade's Window} and the {\it S}-OGLE29 field. We also suggest three further regions of interest, which sample areas of negative $\xi$ away from the minor axis and large latitude; we refer to these as Field A, Field B and Field C. We increase the field size in regions of larger statistical uncertainty, allowing us to sample sufficient number of star particles to provide reasonable predictions. The on-sky positions and sizes of each region of interest are shown in the top panel of  Fig.~\ref{fig:additional-fields}. For each field, we produce \apml\ rotation curves, following the same method as above. The results are presented in the bottom panels of Fig.~\ref{fig:additional-fields}.

The two BTS fields close to the {\it S}-SWEEPS field, {\it S-Stanek's Window} and {\it S-Baade's Window}, only a few degrees apart, have similar $\xi$ values, with anti-symmetric profiles, as in the {\it S}-SWEEPS field. Both {\it S-Stanek's Window} and {\it S-Baade's Window} have \apml\ profiles with increasing amplitude away from the galactic centre, and have similar peak young/old, and a near/far ratio of $\sim 2$.

The {\it S}-OGLE29 field is at the highest latitude of the BTS fields and is further away from the minor axis in a region where $\xi > 0$. This field has a different separation profile: while the rotation curve of the old population remains similar to those previously described, the rotation curve of the young stars crosses the $\apml=0$ line beyond $8\kpc$. The minimum separation between the young and old population occurs in the furthest distance bins. We can understand this behaviour by referring back to the top row Fig.~\ref{fig:XY-grid-sep}. Along longitude $l \approx -7 \degrees$ the nearest distance bins pass through the region where young stars have high positive \apml\ values, where they are streaming at high velocity along the bar edge. The direction of the velocity here is closer to perpendicular to the LOS, which results in the higher \apml\ peak. At the furthest distance bins, we are observing the far end of the bar. The direction of the velocities here are angled more closely parallel to the LOS; therefore, younger stars have a lower \apml\ amplitude, comparable to the value for old stars in the same region. This results in a $\xi$ value larger than the other three BTS fields.

We define Field A to explore the asymmetry in $\xi$. Its location is mirrored across the minor axis from the {\it S}-OGLE29 field, at the same latitude and with the same FOV, within a region where $\xi < 0$. Again the old stars have a rotation curve of increasing \apml\ from the galactic centre. The young stars in this field show little separation in the nearest and furthest distance bins. However, at $7.5 \kpc$, the young population's rotation curve crosses the $\apml = 0$ line and decreases to negative \apml\ more steeply than the older stars, and converges to that of the older population beyond $9\kpc$. At this longitude, the nearest distance bins are within the central bar region, but beyond $\sim 8 \kpc$ the LOS passes through the far edge of the bar, where stars are streaming towards negative longitude. Both the {\it S}-OGLE29 field and Field A demonstrate the effect of observing the proper motions of an angled bar, which we explore further in Section~\ref{s:projections}.

Field B also explores an area where $\xi < 0$ but at larger latitude and longitude. At these higher latitudes, $e_{\xi}$ is larger due to the limited vertical extent of the young population, requiring us to increase the field size considerably. A largely linear profile is seen in both young and old population rotation curves. However, between $8$ and $9 \kpc$ the young population deviates to more negative \apml, similar to Field A. We attribute this deviation to the same effect discussed for Field A, but the separation is smaller here due to being located further away from the minor axis and at higher latitude.

Field C covers a region of high latitude on the minor axis. We see an antisymmetric profile similar to the {\it S}-SWEEPS field; however the young stars have a flatter profile away from the galactic centre. The young stars still have a higher amplitude $\apml$ with a steeper gradient through the galactic centre. The central bins have very few young stars, since the young population is peanut shaped. The $\xi$ value is much larger here due to the effect of perspective.


\section{Projection of Intrinsic Velocities}
\label{s:projections}

\begin{figure*}
	\centerline{
		\includegraphics[angle=0.,width=0.9\hsize]{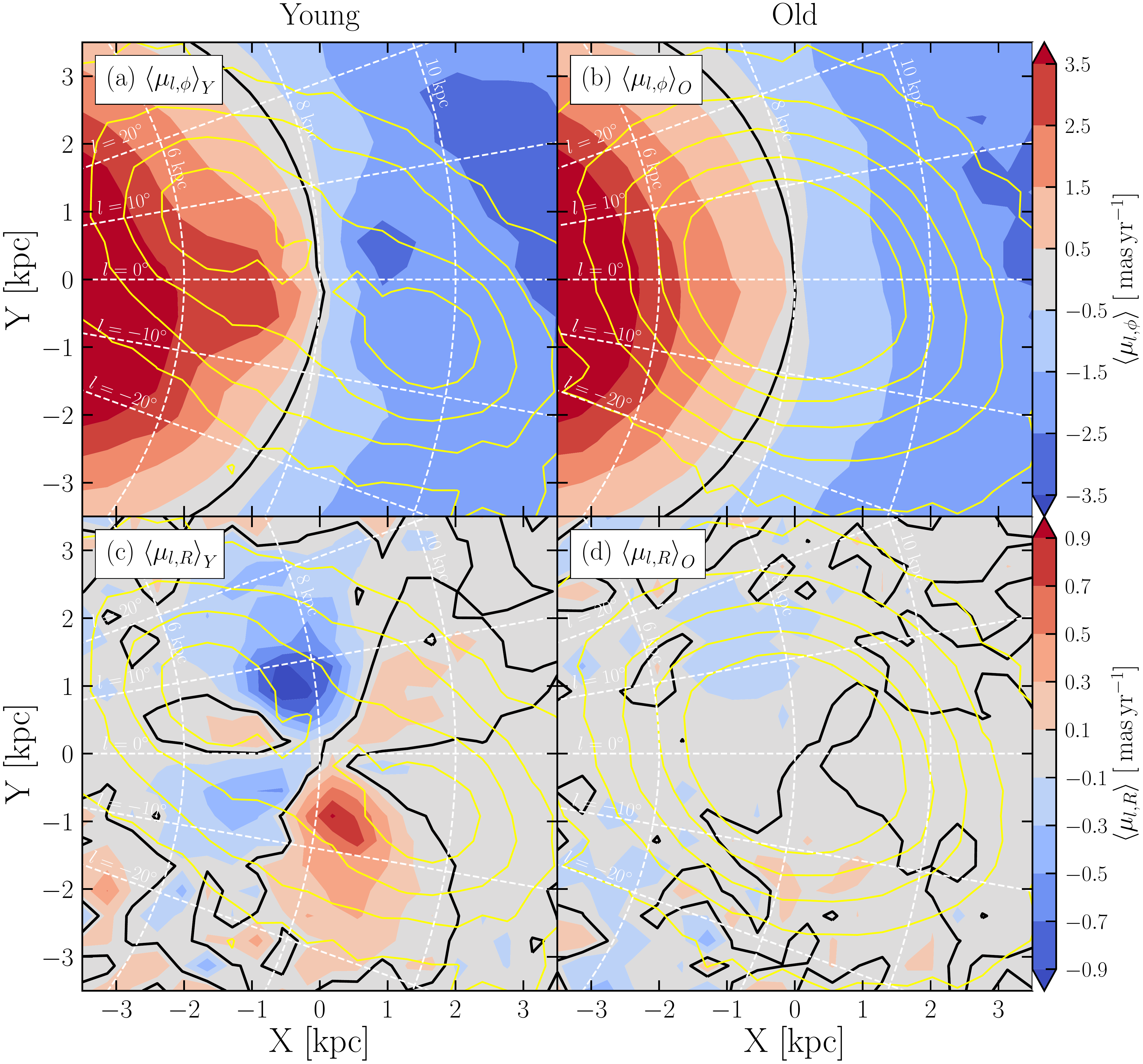}}
	\caption{Projections of the galactocentric intrinsic velocities onto $\mu_l$ in the $(X,Y)$ plane for star particles at $0.5<|z|/\kpc<1.0$. The left-hand column column presents the kinematics of the young population while the right-hand column shows those of the old. The top row shows the projection of galactocentric $\avg{v_\phi}$ onto longitudinal proper motions, the bottom row shows the projection of galactocentric radial velocities $\avg{v_R}$ onto longitudinal proper motions. Yellow contours follow log densities of the corresponding population. Black contours indicate where each velocity component equals zero. White circular dashed lines outline distances $6 \kpc$, $8 \kpc$ and $10 \kpc$, while the white straight dashed lines mark longitudes between $20\degrees$ and $-20\degrees$ in $10\degrees$ intervals. The observer is at $X = -8\kpc$ in this figure.
		\label{fig:XY-cyl-proj}}
\end{figure*}

The rotation curves of the {\it S}-OGLE29 field and Field A demonstrate the clear effect of a non-axisymmetric structure within the bulge region. To illustrate how the galactocentric radial and tangential velocities project onto the observed longitudinal proper motions we now project each galactocentric velocity component individually onto $\hat{l}$, the unit vector in the direction of increasing longitude, \ie\ the tangential direction to the LOS from the Sun. Lines parallel to $\hat{l}$ follow concentric circles centred on the Sun. As these velocity projections are position dependent, not all regions of high-amplitude galactocentric velocity contribute to large proper motions.

We present the projections onto $\hat{l}$ in the $(X,Y)$ plane in Fig.~\ref{fig:XY-cyl-proj}. We denote the $v_R$ and $v_\phi$ projections, respectively, as
\begin{equation}
\mu_{l,R} = \alpha \frac{v_R}{D} \frac{\sin(\phi-l)}{\cos b},
\label{eq:vr-projection}
\end{equation}
\begin{equation}
\mu_{l,\phi} = \alpha \frac{v_\phi}{D} \frac{\cos(\phi-l)}{\cos b},
\label{eq:vphi-projection}
\end{equation}
where $\alpha \approx 0.210 {\>\rm kpc \, s\, km^{-1}}$ and $\phi$ is a star's cylindrical polar angle in the galactocentric frame.

Unsurprisingly, $\avg{v_\phi}$ contributes to \apml\ of both the young and old populations as seen in Panels (a) and (b) of Fig.~\ref{fig:XY-cyl-proj}. In an axisymmetric disc, this would be the only contribution to \apml\ because then $\avg{v_R} = 0$. The $\avg{v_\phi}$ contribution to the old population's proper motions, shown in Fig.~\ref{fig:XY-cyl-proj}(b), has a distribution not much different from that of an axisymmetric disc. Conversely, the young population has stronger rotation closer to the galactic centre, with pronounced twists in the $\avg{\mu_{l,\phi}}$ contours. The regions of low $\avg{v_\phi}$ manifest in the young stars' $\avg{\mu_{l,\phi}}$ distribution as deviations of the velocity contours from being parallel to $\hat{l}$.
In an axisymmetric system, the general trend of increasing velocity dispersion of stellar populations with age would give rise to a separation of the rotation curves purely from $\avg{v_\phi}$, with no contribution from $\avg{v_R}$. However, a stationary axisymmetric system cannot produce a non-zero $\avg{\mu_{l,R}}$. Instead, a bar produces a quadrupolar $\avg{v_R}$ distribution, and hence peaks in $\avg{\mu_{l,R}}$ as seen in the bottom left-hand panel of Fig.~\ref{fig:XY-cyl-proj}. Moreover, the orientation of the MW's bar is such that two of the regions of large $\avg{v_R}$ project almost perfectly into the $\hat{l}$ direction, at positive longitude on the near side and at negative longitude on the far side of the galactic centre. In these regions, the observed longitudinal proper motion has a strong contribution from $\avg{v_R}$. The other two high amplitude $\avg{v_R}$ regions lie at $|l| \lesssim 2 \degrees$ and therefore $\avg{v_R}$ in these regions projects only a small component in the $\hat{l}$ direction. Comparing Panels (a) and (c) of Fig.~\ref{fig:XY-cyl-proj}, it is evident that the main peaks in $\avg{\mu_{l,R}}$ contribute to the total \apml\ with opposite sign to the $\avg{\mu_{l,\phi}}$ for the young population. For example, the peak of negative $\avg{v_R}_Y$ centred near $(X,Y) = (-0.5,0.5)\kpc$ in Panel (c) is within a region of positive $\avg{v_\phi}_Y$ in Panel (a).
The old population has everywhere relatively low $\avg{v_R}$; therefore, its longitudinal proper motion is everywhere dominated by $\avg{v_\phi}$. The young population, having a strong bar, has a strong quadrupolar $\avg{v_R}$.
Consequently, the effect of the bar will be most evident in the kinematics of the young population in the regions where $|\avg{v_R}|$ peaks.

\begin{figure*}
	\centerline{
		\includegraphics[angle=0.,width=.8\hsize]{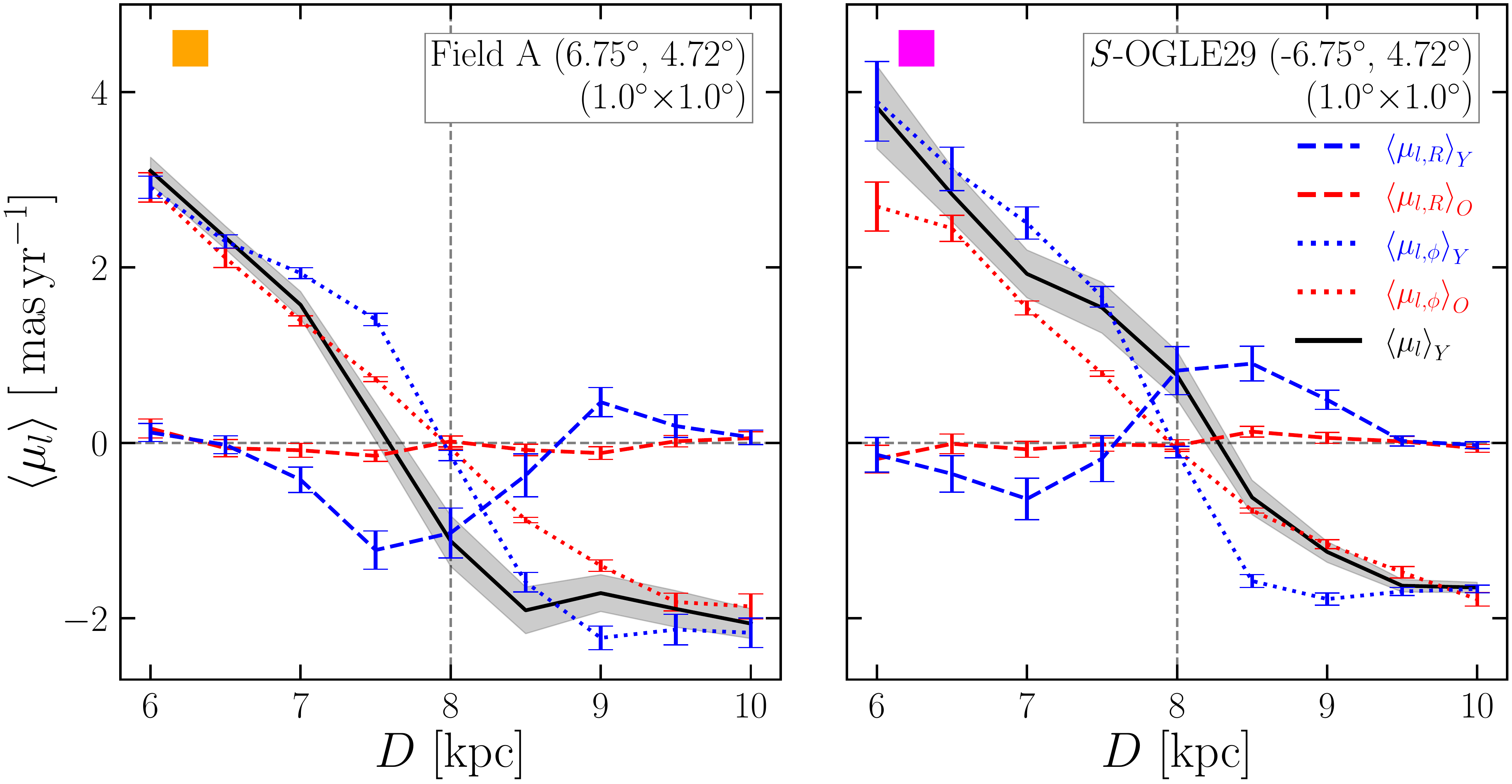}}
	\caption{The contributions of galactocentric $\avg{v_R}$ (dashed lines) and $\avg{v_\phi}$ (dotted lines) to the average longitudinal proper motion rotation curves for young (blue lines) and old stars (red lines) for the simulated Field A (left-hand panel) and the OGLE29 field (right-hand panel). We plot the sum of the two young components, the total observed $\apml_{Y}$ as a black line. The old population has no substantial $\avg{v_R}$ contribution therefore the $\apml_O$ line would lie on top of the $\avg{\mu_{l,\phi}}_O$ line; we thus do not display it. The coloured squares correspond to the field locations indicated in Fig.~\ref{fig:additional-fields}.
		\label{fig:LOS-cyl-proj}}
\end{figure*}

The competing effects of the $\avg{v_R}$ and $\avg{v_\phi}$ contributions to the \apml\ rotation curves give rise to interesting behaviours. We now demonstrate how these two components project onto the rotation curves of the {\it S}-OGLE29 field and Field A. In the left-hand panel of Fig.~\ref{fig:LOS-cyl-proj} we can see that the $\avg{\mu_{l,\phi}}$ component of both populations in Field A follow an antisymmetric profile. However, for the young stars, $\avg{\mu_{l,R}}_Y$ provides a substantial negative contribution slightly short of $8 \kpc$. This contribution acts in opposition to the positive contribution from $\avg{\mu_{l,\phi}}$; as a result, the total $\apml_Y$ for Field A (black line) crosses $\apml = 0$ at $\sim 7.5 \kpc$. Therefore, the radial contribution leads to a sign reversal in the proper motions of young stars and a rotation curve with `forbidden' velocities. We term non-zero velocities at the galactic centre as forbidden because they would not be present in an axisymmetric system. Our usage echos the use of the term for describing gas kinematics at the Galactic Centre \citep[\eg ][]{weiner_sellwood_99}

The young stars in the {\it S}-OGLE29 field also have a rotation curve with forbidden velocities but with a sign reversal from negative to positive. The radial contribution comes somewhat beyond the galactic centre, where the $\avg{v_\phi}$ component is negative, and the $\avg{v_R}$ velocities are positive. The total $\apml_Y$ rotation curve crosses $\apml = 0$ at $\sim 8.25 \kpc$.

The young stars in both of these fields reverse the sign of their proper motions due to the contribution of the radial velocity. The age dependence of bar strength and their resulting velocity profiles demonstrated above are a prediction of kinematic fractionation, where younger populations with lower initial in-plane random motions are less vertically heated, form a stronger bar and a more peanut-shaped bulge \citep{debattista+17}.


\subsection{Quantifying the effect of kinematic fractionation}
\label{ss:deltamu}

\begin{figure*}
	\centerline{
		\includegraphics[angle=0.,width=.95\hsize]{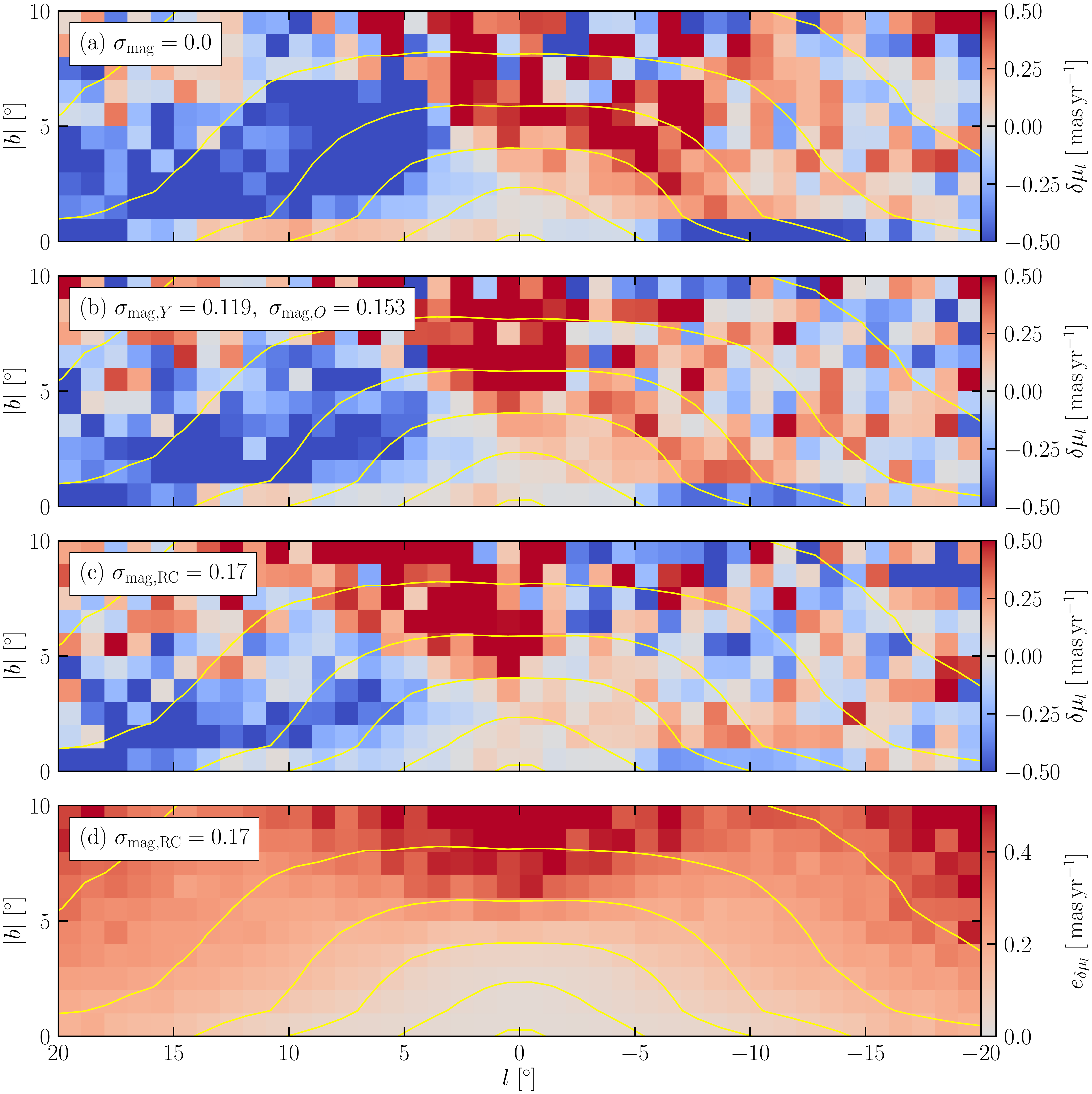}}
	\caption{Top panel: the $\delta \mu_l$ distribution in the bulge region defined as the difference in \apml\ between the young and old populations at $\sim 8 \kpc$. Middle top panel: the same as above but with young and old stars apparent magnitudes convolved with \citetalias{clarkson+18} uncertainties of $\sigma_{\rm mag, Y} = 0.119$ and $\sigma_{\rm mag, O} = 0.153$. Middle bottom panel: the same as above but with both populations convolved with the width of the RC, $\sigma_{\rm mag, RC} = 0.17$. Bottom panel: the calculated error for each field when applying the RC magnitude uncertainties.
		\label{fig:deltav}}
\end{figure*}

We now develop a second metric to quantify the signature of kinematic fractionation within the bulge which is less reliant on deep observations with highly accurate distance determinations. We define a large spatial bin located at $D = 8 \kpc$ with a width of $1 \kpc$, allowing for larger distance uncertainties, then calculate the difference in \apml\ between the young and old populations, \ie\:
\begin{equation}
\delta \mu_l  = \apml_{Y,8 \kpc} - \apml_{O,8 \kpc}.
\label{eq:deltapml}
\end{equation}
Large positive values of $\delta \mu_l$ correspond to rotation curves where the young population have larger positive \apml\ than the old population within this central bin, which are forbidden velocities at negative longitude. We use $\delta \mu_l$ to measure forbidden velocities in the bulge whilst also taking into account the expected observational distance uncertainties. Following from our previous analysis, we expect to measure positive $\delta \mu_l$ values in the direction of the OGLE29 field and negative values in the direction of Field A. We assume every star in the model is an RC star with absolute magnitude $M_K=-1.61$, and calculate their apparent magnitudes as we did in Section~\ref{ss:milkyway} to more closely approximate observations. We assume extinction to be uniform across the bulge region for simplicity. Reproducing this work observationally would rely on extinction corrections being made for the tracer populations used.
We then define the magnitude range equivalent to $8 \pm 0.5 \kpc$, which is $12.75 - 13.05 \mags$, a bin width of $0.3 \mags$. We present $\delta \mu_l$ in the $(l,|b|)$ plane in Fig.~\ref{fig:deltav}, under three different assumptions for the distance uncertainty, $\sigma_{\rm mag}$: no uncertainty, SWEEPS field uncertainties and RC uncertainties. For the SWEEPS uncertainties, we assume the metal-rich and metal-poor magnitude uncertainties from \citetalias{clarkson+18} apply to our young and old populations respectively, $\sigma_{\rm mag, Y} = 0.119$ and $\sigma_{\rm mag, O} = 0.153$. For the RC uncertainties, we apply the width of the RC distribution, as in Section~\ref{ss:milkyway}, $\sigma_{\rm mag, RC} = 0.17$ to both populations. To estimate the uncertainty in $\delta \mu_l$, we add in quadrature the \apml\ uncertainty for the young and old populations.

In the top panel of Fig.~\ref{fig:deltav}, we present $\delta \mu_l$ across the bulge assuming no magnitude uncertainties. The distribution of $\delta \mu_l$ has a left/right asymmetry with negative values for fields $l > 0 \degrees$ and positive values at $l < 0 \degrees$. The peaks in the amplitude of $\delta \mu_l$ occur around $(|l|,|b|) = (6\degrees,5\degrees)$. Along the minor, axis we expect $\delta \mu_l \sim 0$ as there is only a small contribution from galactocentric radial velocities in this region. Away from the minor axis, close to the locations of the OGLE29 field and Field A, we find large values $|\delta \mu_l|$. Regions with $\delta \mu_l > 0$ are present near the OGLE29 field at negative longitude, while regions of $\delta \mu_l < 0$ surround a large area around Field A, at positive longitude. Regions of positive $\delta \mu_l$ at high latitude on the minor axis are due to the bin centre being at a cylindrical radius closer than the galactic centre at this latitude, worsened by large bin width used.

With increasing distance uncertainties, as in the two middle panels of Fig.~\ref{fig:deltav}, the peaks near $(|l|,|b|) = (6\degrees,5\degrees)$ become weaker, with smaller $\delta \mu_l$ values in general. The overall distributions still retain a left/right asymmetry but with more fields having small $\delta \mu_l$ indistinguishable from $\delta \mu_l = 0$. In all panels, the negative $\delta \mu_l$ region is larger and has higher amplitude than the positive one, due to the bar's orientation. In the direction of the OGLE29 field and Field A typical values of $|\delta \mu_l| > 0.5 \masyr$ at $8 \kpc$ correspond to $\approx 20 \kms$ difference in heliocentric tangential velocities.

We present the $e_{\delta \mu_l}$ map for the RC magnitude uncertainties in the bottom panel of Fig.~\ref{fig:deltav}. The variation in $e_{\delta \mu_l}$ between the three levels of magnitude uncertainties is minimal and retains the general trends. Similar to $e_{\xi}$, the $e_{\delta \mu_l}$ distribution is peanut shaped with the region of the largest uncertainty at high latitude on the minor axis, again because of the lower number of young stars there.

The map of $\delta \mu_l$ is in good agreement with the map of $\xi$, in as much as large separations in velocities are observed in regions away from the minor axis. We have demonstrated here that large amplitude values of $\delta \mu_l$ are a result of rotation curves with forbidden velocities (see Fig.~\ref{fig:LOS-cyl-proj}), which are the result of radial velocity contributions from bar supporting orbits.


\section{Comparison With a Weaker B/P Model}
\label{s:m2-comp}

We briefly explore a second model (hereafter Model 2) which forms a bar later in its evolution; the bar is weaker and produces a weaker B/P bulge than the fiducial model. The initial conditions from Section~\ref{s:simulation} remain the same; however, Model 2 has different subgrid physics and forms a bar of length $\sim 2.5 \kpc$ between $4-6 \Gyr$ (versus $\sim 3 \kpc$, forming at $2-4 \Gyr$). To directly compare with the fiducial model, we scale and align Model 2 following the same procedures outlined in Section~\ref{ss:milkyway} but with a spatial scaling factor of 2 instead of 1.7. Owing to the stochasticity inherent in bar evolution \citep{sellwood_debattista09}, the two bars do not evolve in the same way. The bar length at $10 \Gyr$ in Model 2 after scaling is $r_{bar} = 4.80~\pm~0.90 \kpc$, and the double RC appears only weakly at $|b| = 6 \degrees$ as a result of its more limited B/P growth compared to the fiducial model. We plot the radial profiles of the $m=2$ Fourier moment amplitude and phase along with the evolution of the global bar amplitude for both models in Appendix A. The cumulative distribution of ages within Model 2's bulge reveals that it has a lower star formation rate at the beginning of the simulation; our cut of old stars at age $> 9 \Gyr$ therefore represents a smaller fraction of the bulge population. The selection of young stars (age $< 7 \Gyr$) samples a higher fraction of stars born before and during the formation of the bar (85\%), which lowers the overall bar strength of this population.

We measure our metrics of kinematic separation, $\xi$ and $\delta \mu_l$, across the bulge of Model 2, which we present in Appendix A. Our measurements show the same global trends as in the fiducial model. The map of $\xi$ is asymmetric about the minor axis with $\xi < 0$ at positive longitudes and $\xi > 0$ at negative longitudes. The amplitudes of $\xi$ are lower than in the fiducial model with a steeper decreasing gradient with increasing latitude. Model 2 has similar values of the uncertainty, $e_{\xi}$, however, its distribution is less peanut shaped.
The map of $\delta \mu_l$ also matches the general trends of the fiducial model's with a left-right asymmetry, $\delta \mu_l < 0$ at positive longitude and $\delta \mu_l > 0$ at negative longitude but also has generally lower amplitude values. The $e_{\delta \mu_l}$ distribution is also more box shaped.

\begin{figure*}
	\centerline{
		\includegraphics[angle=0.,width=0.8\hsize]{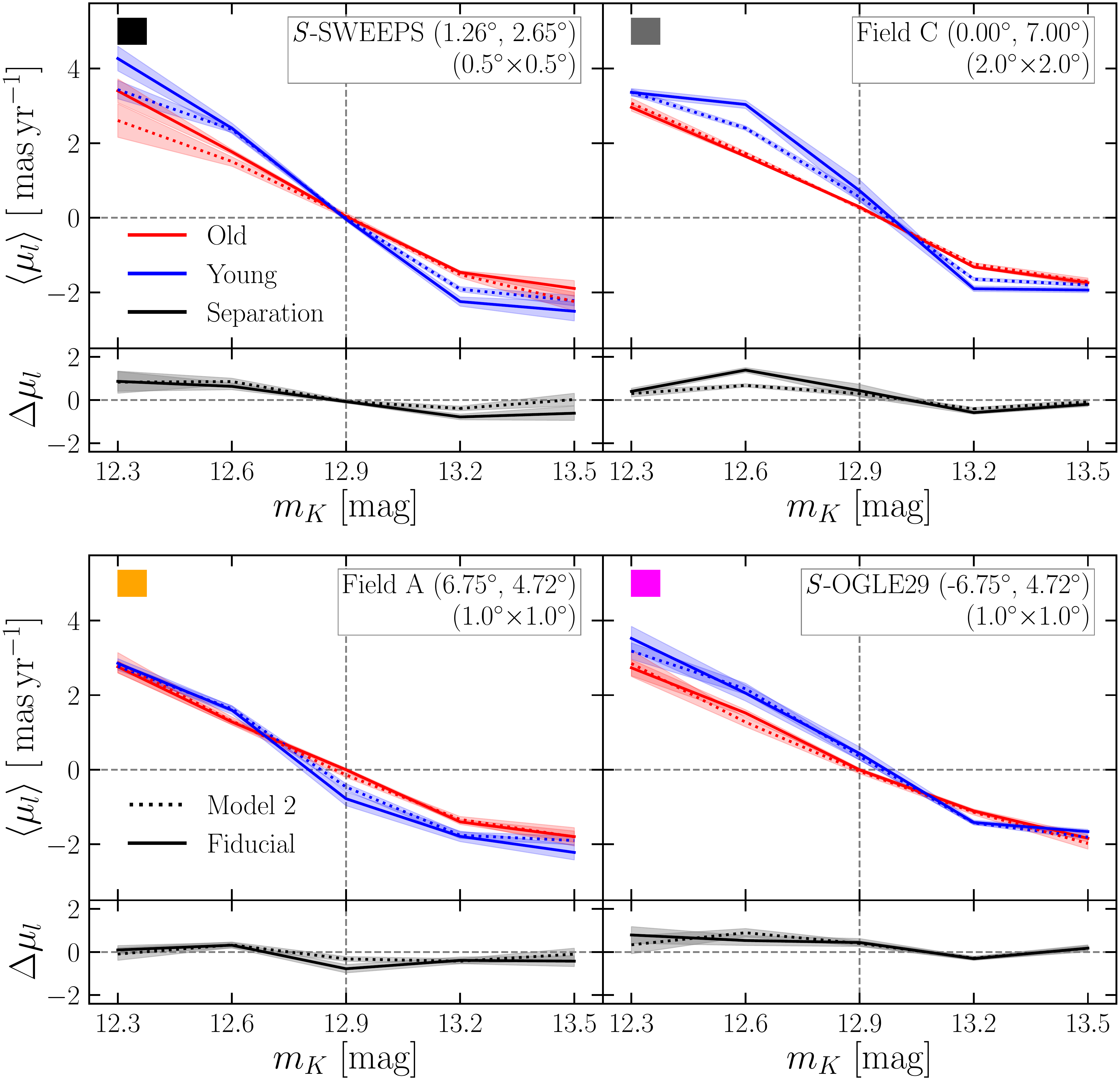}}
	\caption{Average longitudinal proper motion rotation curve for young and old stars, and the separation between them using magnitude bins for four key fields in the bulge. The rotation curves and separation from the fiducial model are plotted as dotted lines whereas Model 2 is plotted as solid lines. The field name, location, and FOV are labelled in the top right-hand of each panel. The coloured squares correspond to the field locations indicated in Fig.~\ref{fig:additional-fields}.
		\label{fig:LOS-model2}}
\end{figure*}

We present a comparison of each model's rotation curves for the simulated {\it S}-SWEEPS and {\it S}-OGLE29 fields along with Field A and Field C in Fig.~\ref{fig:LOS-model2}, where now we have used magnitude bins of $0.3 \mags$. instead of distance. We use the magnitude range equivalent to span $5.75-10.25 \kpc$ to present predictions of $\xi$ and $\delta \mu_l$ simultaneously for these key fields (\ie\ $\delta \mu_l$ is the \apml\ difference in the central bin at $12.9 \mags.$). This also allows us to test our metrics using a methodology closer to MW observations. $\xi$ and $\delta \mu_l$ for these fields are shown in Table \ref{tab:xicomparison}.

\begin{table}
    \begin{centering}
        \begin{tabular}{l|l|c|c|c|c}\hline
        \multicolumn{1}{c}{Field} &
        \multicolumn{1}{c}{Model} &
        \multicolumn{1}{c}{$\xi$} &
        \multicolumn{1}{c}{$e_\xi$}
        &
        \multicolumn{1}{c}{$\delta \mu_l$}
        &
        \multicolumn{1}{c}{$e_{\delta \mu_l}$}\\ \hline
        {\it S}-SWEEPS & Fiducial & 0.005 & 0.091 & -0.073 & 0.068  \\
        {\it S}-SWEEPS & Model 2 & 0.188 & 0.093 & -0.065 & 0.084 \\
        Field C  &  Fiducial & 0.217 & 0.055 & 0.435 & 0.293  \\
        Field C &  Model 2 & 0.117 & 0.037 & 0.295 & 0.118 \\
        Field A  &  Fiducial & -0.177 & 0.064 & -0.776 & 0.192 \\
        Field A &  Model 2 & -0.092 & 0.068 & -0.322 & 0.111 \\
        {\it S}-OGLE29  &  Fiducial & 0.245 & 0.076 & 0.439 & 0.183  \\
        {\it S}-OGLE29 &  Model 2 & 0.226 & 0.076 & 0.398 & 0.117 \\ \hline
        \end{tabular}
    \caption{Calculated values of $\xi$ and $e_\xi$ (with units of $\rm kpc \cdot \rm mas\, yr^{-1}$), $\delta \mu_l$, and $e_{\delta \mu_l}$ (with units of $\rm mas\, yr^{-1}$) for the simulated {\it S}-SWEEPS, {\it S}-OGLE29, Field A, and Field C in the fiducial model and Model 2.}
    \label{tab:xicomparison}
    \end{centering}
\end{table}

Although now binning in magnitude, the rotation curve profiles in both models qualitatively match those presented in Fig.~\ref{fig:additional-fields}. Here, we discuss only the differences between the two models.

In the {\it S}-SWEEPS fields, the amplitudes of \apml\ are lower for both the young and old populations in Model 2 than the fiducial model. However, considering the separation profile, Model 2 has similar separation on the near side but has weaker separation beyond the galactic centre resulting in a larger $\xi$ value. As expected for a central longitude, the {\it S}-SWEEPS field has very low $\delta \mu_l$ in both models.

Field C in Model 2 also has lower \apml\ amplitudes. The old population rotation curves in the two models overlap each other. However, there is a large difference between the rotation curves of the young populations in the two models: in the fiducial model, the rotation curve is steeper at the galactic centre. Model 2 also has lower separation in the nearest bins resulting in a lower $\xi$. Field C is at a large latitude on the minor axis and as a result there is a small perspective effect contributing to $\delta \mu_l$. At this latitude, the central ($8 \kpc$) bin is located in front of the galactic centre by $\sim 0.16 \kpc$, which results in both populations crossing the $\apml = 0$ line beyond the central bin, and thus a positive $\delta \mu_l$ results. This effect is also seen in the bottom right-hand panel of Fig.~\ref{fig:additional-fields}.

Field A in both models has similar rotation curves with the main difference being in the degree of deviation into forbidden velocities. Model 2 has lower separation at the central bin resulting in lower $\xi$ and $\delta \mu_l$ values. The two models differ very little in the {\it S}-OGLE29 field, with a slightly larger $\delta \mu_l$ in the fiducial model.

The combined effects of later bar formation and the weaker bar and B/P in Model 2 result in longitudinal proper motion rotation curves that are qualitatively similar in profile but have lower separations and forbidden velocities as measured by both the $\xi$ and the $\delta \mu_l$ metrics. Thus, the separation amplitudes and the global trends of $\xi$ and $\delta \mu_l$ provide important information on the relative bar strength of each population and may be useful in constraining the MW's bar and bulge.


\section{Discussion}
\label{s:discussion}

Kinematic differences between different populations in the MW's bulge have been explored in many studies. Metallicity is most often used to separate bulge stellar populations. Metal-rich bulge stars typically have higher radial velocity dispersion than metal-poor stars \citep{rich_90, sharples+90, minniti_96} and the metal-poor population has a shallower velocity dispersion gradient with latitude \citep{ness+13a, ness+13b, ness+16b, zoccali+17}. Observations at Baade's Window have shown that metal-rich stars have non-zero vertex deviation, whereas that of the metal-poor stars is nearly zero \citep{soto+07, babusiaux+10, hill+11, vasquez+13}.
Even amongst the (old) RR~Lyrae (RRL), metallicity separates different populations. \citet{du+2020} and \citet{kunder+2020} found that metal-rich RRL trace a (weak) bar and have angular velocities slightly larger than metal-poor RRL, which do not trace any bar.
Populations can also be separated by their $\alpha$-abundance. \citet{queiroz+2020b} explored the chemo-dynamics of the bulge using APOGEE and \gaia\ data. They found two distinct components when considering the $v_\phi$ vs Galactocentric radius distribution. One component is a low-$\alpha$ population with high rotational velocities and the other has high-$\alpha$ concentrated at small radii and with near-zero or negative $v_\phi$. However, the single chemical track of the bulge implies that separating populations by the $\alpha-$abundance is similar to separating by metallicity.

In summary, these observational results point to stronger barred streaming motions in metal-rich stars, which is borne out by models and simulations \citep{portail+17, debattista+20}.
However, the origin of each component within the bulge is still a matter of debate. Whilst it is possible for (part of) the metal-poor component to be a classical bulge formed through mergers, most of it may also have formed \textit{in situ}.
Indeed, using the same fiducial model as here, \citet{debattista+17} showed that the velocity dispersions of stars separated by age qualitatively match the above trends in the ARGOS data provided a halo-like population was added to the very oldest, most metal-poor stars.
Observationally stars with metallicity $\feh < -1$ represent only $\sim 5 \%$ of all bulge stars \citep{ness_freeman_16}, with no more than $1/3$ of stars with metallicity $\feh \leq -0.8$ potentially being an accreted population \citep{horta+2020}. Recent studies of zoom-in simulations of MW-like galaxies by \cite{fragkoudi+2020} suggest that the bulge contains a negligible fraction of accreted stars \cite[see also][]{buck+19}. Isolated simulations also showed that such a component could not be larger than $\sim 8 \%$ of the disc mass \citep{jshen+10}.

An efficient way to probe the kinematics of bulge stellar populations as a function of distance was presented by \citetalias{clarkson+18} for a sample of just under ten thousand main-sequence stars in a deep {\it HST} field combining SWEEPS and BTS data. Separating these by relative photometric metallicity, \citetalias{clarkson+18} produced longitudinal proper motion rotation curves of `metal-rich' and `metal-poor' samples. They found that metal-rich stars have larger amplitude longitudinal proper motions. In this paper, we have simulated the SWEEPS+BTS field ({\it S}-SWEEPS) using an isolated, star-forming model scaled to approximate the MW. Our young and old populations match the trends of the MW metal-rich and metal-poor main-sequence stars in as much as the young (metal-rich) stars having a larger amplitude \apml\ along the LOS than the old (metal-poor) ones. The amplitude between the young and old populations differs by roughly a factor of 2, in good agreement with the observations of \citetalias{clarkson+18}. Thus we conclude that the trends in the rotation curves of the bulge can be reproduced without the need for an accreted population.

To help prepare for future studies, we have quantified the difference between the rotation curves of the two populations by defining a separation amplitude, $\xi$, as the sum of the difference between averages of longitudinal proper motion in distance bins along an LOS. We have demonstrated, using Monte Carlo resampling to account for distance and velocity uncertainties, that $\xi$ is similar between our model and the SWEEPS+BTS \citetalias{clarkson+18} data, despite the differences in sample selection between the model and observations. We have measured $\xi$ across the entire bulge region covering $|l|\leq20\degrees$, $2\degrees\leq|b|\leq10\degrees$, and $5.75 \leq D/\kpc \leq 10.25$. Both the distribution of $\xi$ within the bulge and the rotation curves of key fields indicate that the rotation curve profiles change with longitude and latitude. We interpret these variations as differences in the intrinsic velocity distributions of the two populations.

The galactocentric cylindrical velocities of (relatively) young stars match the expected signature of stars on strongly barred orbits; in contrast, the old stars trace a weaker bar. Here we have selected stars based on their age. While not a perfect match for metallicity, our results suggest that the bar should be more metal-rich than the rest of the bulge population. The different velocity profiles reflect the underlying density distributions and relative bar strengths of the populations. Recent studies have indeed shown that the bar is more metal-rich than the off-bar regions \citep{wegg+19, queiroz+2020b} \citep[but see][for a different view]{bovy+19}.

We have studied how the intrinsic velocities of stars on bar orbits project onto longitudinal proper motions by considering the radial as well as tangential velocity components separately in both galactocentric and heliocentric coordinates. We find, in the young population, regions of high galactocentric radial velocities in the $(X, Y)$ plane as a quadrupole rotated by $\sim 45\degrees$ relative to the bar axes. With the MW bar inclination angle of $\sim 27 \degrees$, two of these regions project onto longitudinal proper motions at lines of sight away from the minor axis ($|l| \approx 6 \degrees$). Fields which intersect these regions have rotation curve profiles quite different to those of the SWEEPS field and other fields on the minor axis. The galactocentric radial velocity contribution is in the opposite direction to the contribution from the galactocentric tangential velocity, resulting in a rotation curve with `forbidden velocities': negative \apml\ at positive longitudes, and positive \apml\ at negative longitudes. The {\it S}-OGLE29 field is one such case, and as a result, the young stars have a rotation curve that changes sign (crosses the $\apml = 0$ line) beyond $8~\kpc$. The old population shows no such deviations as a result of much lower galactocentric radial velocities produced by their weaker bar. Since the forbidden velocities would not be present in an axisymmetric system, they are the best probe of the variation of the bar strength. Thus the minor axis is not the ideal probe of the bar in proper motion rotation curves. We have analysed only two bar models; therefore we defer a deeper quantification of the relationship between intrinsic kinematics and bar strength to a future study.


\subsection{Future prospects}
\label{ss:futurepros}

We have predicted the lines of sight which have rotation curves with large galactocentric radial velocity contributions acting in opposition to galactocentric tangential velocity in the same region, resulting in forbidden velocities.
These regions provide clear indications of kinematic differences due to a stronger bar in the relatively young populations. We consider the difference in velocities between model young and old stars within a magnitude bin equivalent to $1 \kpc$ at $R_\odot = 8 \kpc$, denoted as $\delta \mu_l$, assuming the stars have fixed absolute magnitude. We find regions centred near $(|l|,|b|) = (6\degrees,5\degrees)$ show the largest separation, with higher amplitudes of $|\delta \mu_l|$ at positive longitude (owing to the bar orientation). These would be very fruitful targets for future observations.

The predictions in this study provide a framework for the observational testing of evolutionary pathways of the MW bulge and bar, such as the time of bar formation. While the separation of populations in these models has used stellar ages, a similar separation can be achieved in chemical (\feh) space, as demonstrated in the SWEEPS field by \citetalias{clarkson+18}. The remaining {\it HST}-BTS fields offer an opportunity to test the predictions in Fig.~\ref{fig:additional-fields} and Fig.~\ref{fig:LOS-model2}. Proper motions are publicly available for the remaining fields, and we plan to use the same method adopted by \citetalias{clarkson+18} for the SWEEPS field to determine the photometric metallicity and distances for main-sequence stars in the OGLE29, {\it Stanek's} and {\it Baade's} windows (Clarkson \etal\ {\it in preparation}). The derived rotation curves for populations split by metallicity will be compared to those in this paper for the relatively young and old stars. The comparison of the OGLE29 window would be the most critical test of the results presented here as it has a distinct rotation curve profile with forbidden velocities. Although we find stronger signals at positive longitude in field A (see Fig.~\ref{fig:deltav}), the OGLE29 field presents an opportunity to test our predictions with data already available.

The {\it Nancy Grace Roman Space Telescope} ({\it RST}) promises to provide high-precision astrometry for $\sim100$~million stars within the bulge \citep{WFIRST+19}. A shallow, multiepoch survey with {\it RST} directed at the key fields identified in this paper would be very useful to constrain the bulge/bar rotation curves. On the other hand, an All-Sky near-IR astrometric space mission \cite[GaiaNIR,][]{hobbs+2021} would provide homogeneous proper motions, parallaxes, and NIR magnitudes down to the Main Sequence Turn-Off (MSTO) in regions close to Galactic plane, thus facilitating the study of proper motion rotation curves as presented in this work, as probes for the formation of the bulge, and its dynamical evolution.

Future ground-based spectroscopic surveys (e.g. APOGEE-2, MOONS, 4MOST) \citep{zasowski+17, moons_gonzalez, 4mostbulge} will collect high-resolution spectra for millions of bulge red giant stars, measuring metallicity, elemental abundances, and radial velocities. These large samples, when combined with the 2D motions from extensive photometric surveys such as VVV and \gaia, have the potential to facilitate the investigation of rotation curves of chemically distinct populations. It is thus critical that surveys deliver sufficiently large samples in key lines of sight for the measurement of statistically significant kinematic separations, after decomposing in chemical and distance space.

The Vera C. Rubin Observatory/LSST has the potential to produce a one-of-a-kind synoptic data set to test the predictions presented in this study. In particular, a multiepoch survey of the Galactic bulge region, deep enough to reach the MSTO, would provide the ideal data set to measure both ages and proper motions \citep[][LSST bulge white paper]{gonzalez+18} and apply the methods used here. A key output of LSST data would be a homogeneous, wide-field map of these properties (similar to the map in Fig.~\ref{fig:field-sepamp} and Fig.~\ref{fig:deltav}). This `definitive map' would allow us to characterize the morphologies of different stellar populations of the bulge and bar in unprecedented detail, answering fundamental questions about the formation of the MW bar.


\subsection{Summary}
\label{s:summary}

We summarize our main conclusions as follows:

\begin{enumerate}
    \item We have shown that the longitudinal proper motion rotation curves of old and (relatively) young stars are distinct, with the rotation curves of young stars generally having larger amplitudes. Our results are in agreement with observations of the SWEEPS field within the MW, which showed that the metal-rich population has a higher amplitude proper motion rotation curve than the metal-poor one (\citetalias{clarkson+18}). This result does not require the presence of an accreted population (see Section~\ref{s:separation}).

    \item We have presented maps of the intrinsic kinematics of each population to help understand the observations. The galactocentric cylindrical velocities of young stars are consistent with bar aligned orbits, in contrast to the nearly axisymmetric velocity distributions of old stars, which reflect their respective underlying density distributions. We demonstrate how the intrinsic velocities project onto longitudinal proper motions. Large galactocentric radial velocity contributions (in the young populations) produce rotation curves with forbidden velocities, which would not be present in an axisymmetric system (see Section~\ref{ss:bulgekin}).

    \item We have defined two metrics to quantify the difference between the rotation curves of young and old populations, and predict their variation across the bulge. We show that the rotation curves of young and old populations in fields which intersect the bar away from the minor axis have non-antisymmetric separation profiles. These effects are due to the large galactocentric radial velocities of young stars which, along these lines of sight, project into forbidden proper motions (see Sections \ref{ss:sepamp} and \ref{ss:deltamu}).

    \item We have demonstrated that the rotation curve separations can be explained by the distinct kinematics of populations separated by an evolving bar, as predicted by kinematic fractionation \citep{debattista+17}, without the need for an accreted component. However, rotation curve separation would also naturally be present in an axisymmetric system because of the increasing asymmetric drift with population age. Therefore, it is the longitudes with forbidden velocities, which probe the variation of the bar's strength with age. (See Section~\ref{s:projections}.)

    \item Finally, we present predictions of our two metrics and the profiles of rotation curves for key fields within the MW Bulge (see Section~\ref{s:m2-comp}).
    These will allow for follow-up study with {\it HST}-BTS data (Clarkson \etal\ {\it in preparation}) along with future survey missions such as {\it RST} and LSST. We recommend deep observations of fields away from the minor axis, close to the regions of $(|l|,|b|) = (6\degrees,5\degrees)$ where we have demonstrated rotation curves have forbidden velocities.

\end{enumerate}


\section*{Acknowledgements}

VPD is supported by STFC Consolidated grant \#~ST/R000786/1. SGK is supported by the Moses Holden Studentship, with particular thanks to Patrick Holden. The simulations used in this paper were run at the High Performance Computing Facility of the University of Central Lancashire. This work is based partially on observations made with the NASA/ESA \textit{Hubble Space Telescope (HST)} and obtained from the data archive at the Space Telescope Science Institute (STScI). The \textit{HST} comparisons reported in this paper rely on products from \textit{HST} Guest Observer programs 9750, 12586 and 13057 (P.I. Sahu), 12020 (P.I. Clarkson), 11664 and 12666 (P.I. Brown), support for which was provided by NASA through grants from STScI. STScI is operated by the Association of Universities for Research in Astronomy, Inc., under NASA contract NAS 5-26555 The authors are grateful for the anonymous referee's comments which helped to clarify and improve this paper. The analysis in this paper made use of the \texttt{PYTHON} packages  \texttt{NUMPY, SCIPY, PANDAS, GALPY, PYNBODY} and \texttt{JUPYTER} \citep{numpy, scipy, pandas, galpy,pynbody, jupyternotebooks}. Figures in this work were produced using the \texttt{PYTHON} package \texttt{MATPLOTLIB} \citep{matplotlib}.


\section*{Data availability}

The data underlying this article will be shared on reasonable request to the corresponding author.




\bibliographystyle{mnras}
\bibliography{PMBS.bib}



%
\appendix
\section{Separation of Kinematics in Model 2}

Here, we present the properties of Model 2, a simulation with the same initial conditions to our fiducial model but different subgrid physics to those outlined in Section~\ref{ss:milkyway}. The following figures compare the properties of the bar and bulge populations of Model 2 with those of the fiducial model. We also present the equivalent maps of our $\xi$ and $\delta\pml$ metrics for Model 2.

\begin{figure}
	\centerline{
		\includegraphics[angle=0.,width=\hsize]{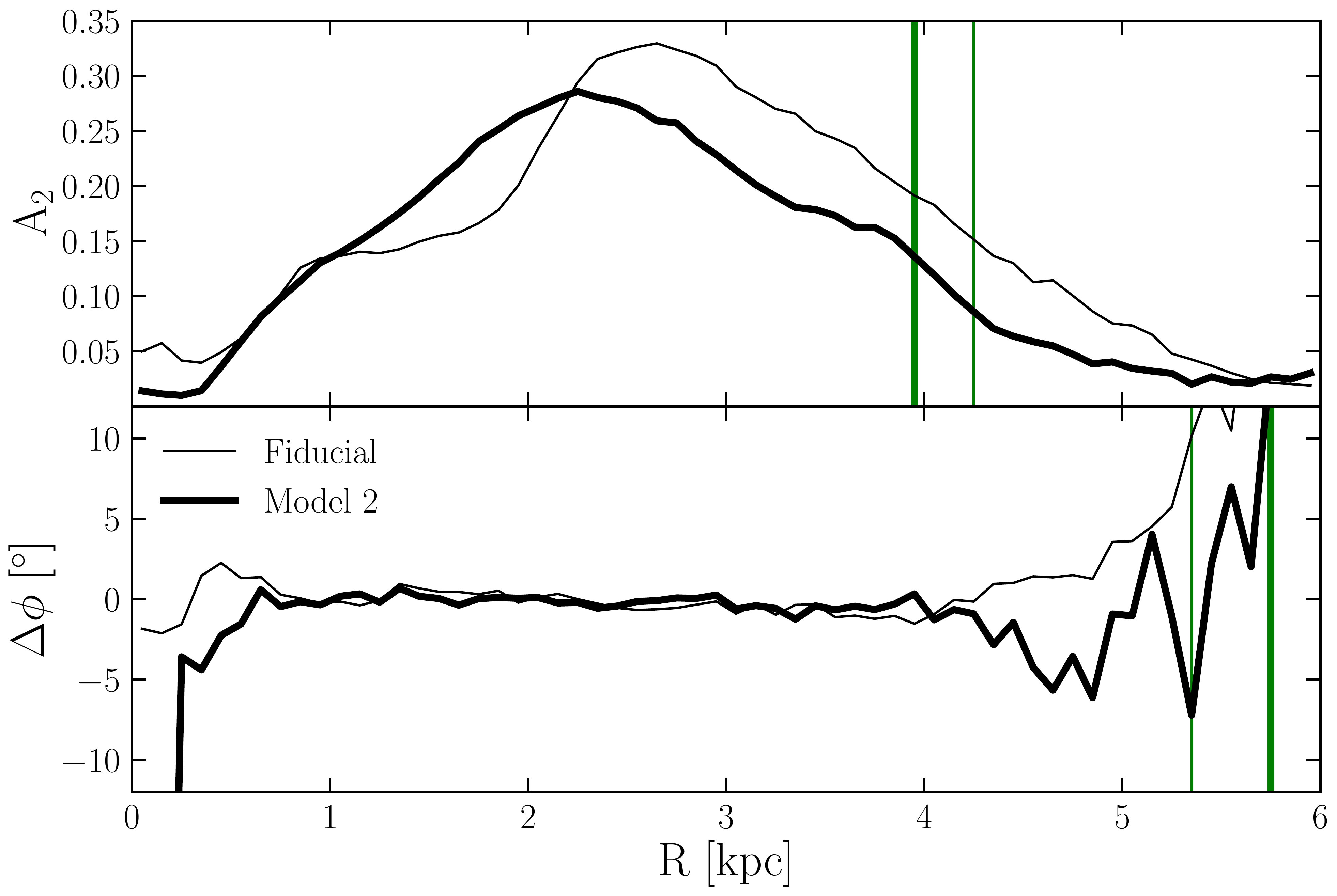}}
	\caption{Top panel: the radial profile of the $A_2$ Fourier amplitude at time $t=10 \Gyr$ of the fiducial model and Model 2. Bottom panel: the change in phase angle of the $m=2$ mode with radius at $t=10 \Gyr$. Vertical green lines indicates where $A_2$ reaches its half maximum value and $|\Delta \phi| >10 \degrees$ for each model. Averaging these two values results in bar radial extents of $4.85\pm0.55\kpc$ and $4.80\pm0.90\kpc$ for the rescaled fiducial model and Model 2 respectively.
		\label{fig:bar-amp-r}}
\end{figure}

\begin{figure}
	\centerline{
		\includegraphics[angle=0.,width=\hsize]{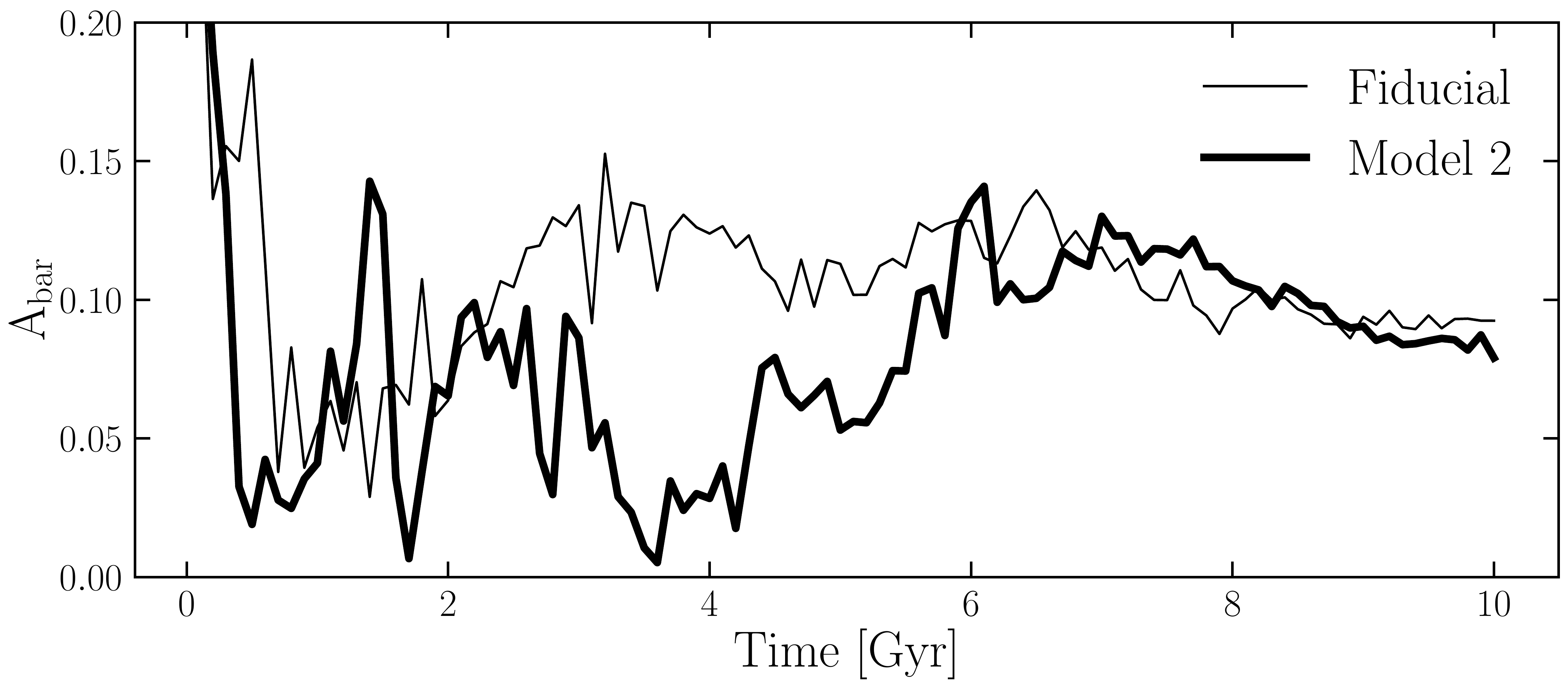}}
	\caption{The global bar amplitudes of the fiducial model and Model 2 versus time. The major growth period for the fiducial model is between $2$ and $4 \Gyr$, and $4$ and $6 \Gyr$ for Model 2.
		\label{fig:bar-amp-time}}
\end{figure}

\begin{figure}
	\centerline{
		\includegraphics[angle=0.,width=\hsize]{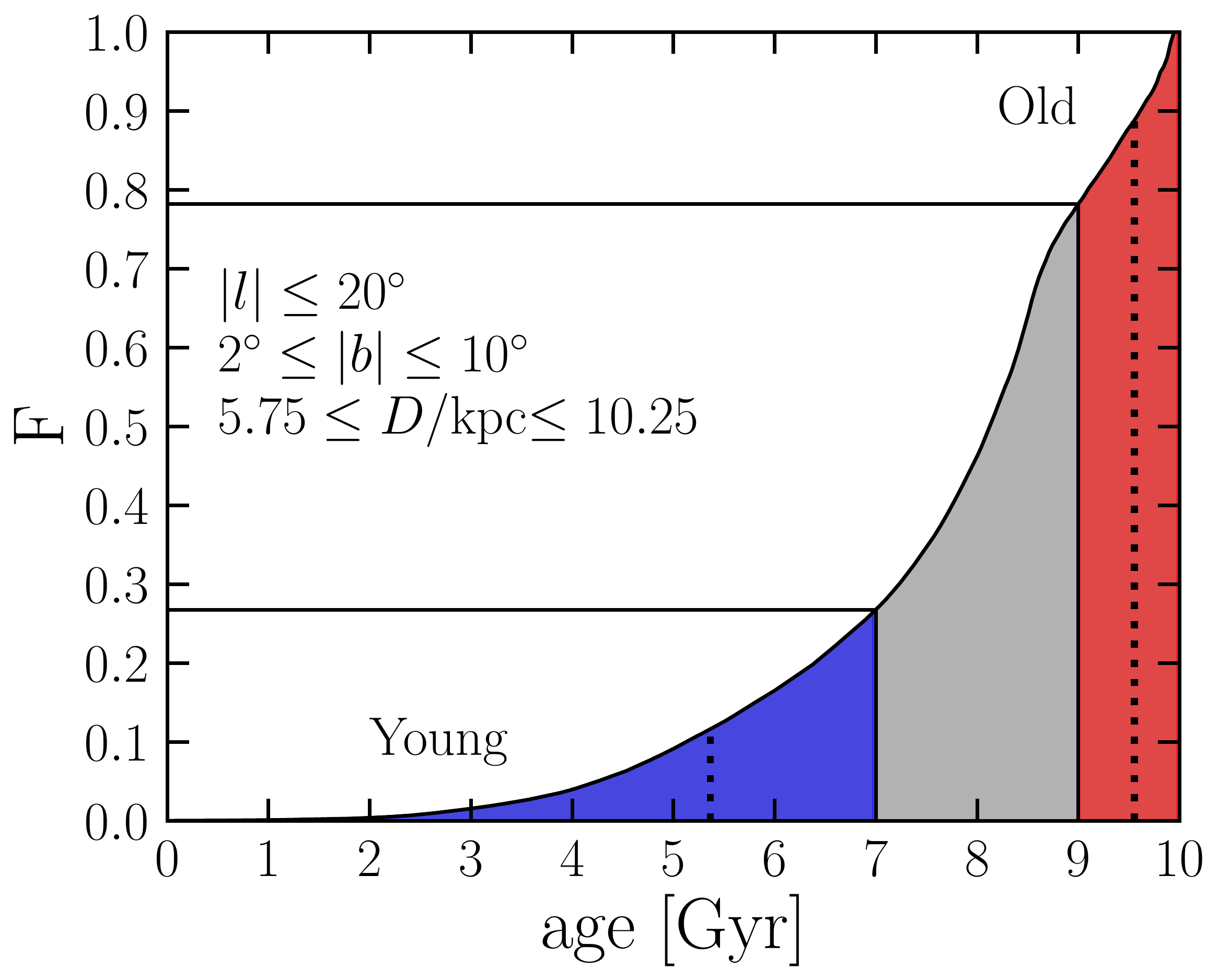}}
	\caption{The cumulative distribution of ages within Model 2's bulge region, defined at top left-hand side, and our definition of the young (blue) and old population (red). The average age for the two populations (vertical black dashed lines) is $5.4$ and $9.6\Gyr$, respectively.
		\label{fig:agedistmodel2}}
\end{figure}

\begin{figure}
	\centerline{
		\includegraphics[angle=0.,width=\hsize]{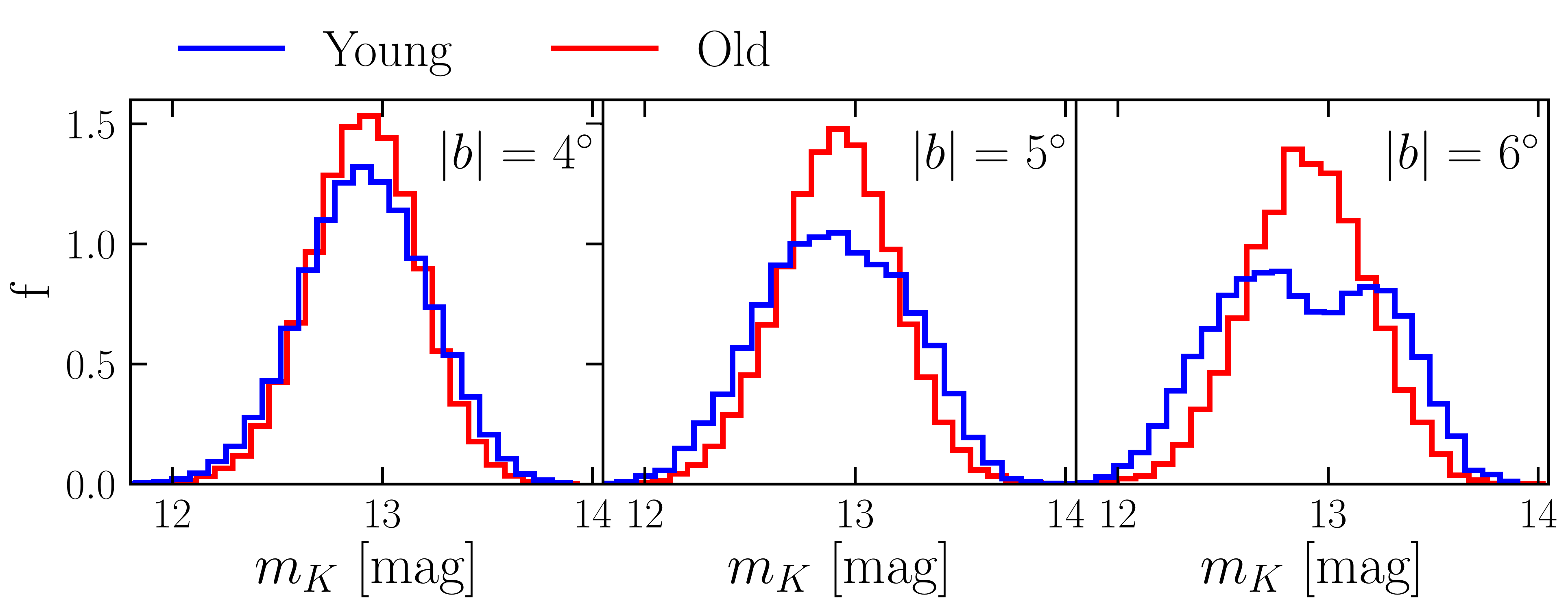}}
	\caption{Unextincted apparent magnitude distributions of simulated RC stars along the LOS within $|l|<4\degrees$ for $|b|=4\degrees$ (left-hand panel), $5\degrees$ (middle panel) and $6\degrees$ (right-hand panel) with $\delta |b| = 0.25\degrees$ in Model 2. Young (age $<7\Gyr$) and old (age $>9\Gyr$) stars are represented by the blue and red histograms, respectively.  The magnitude distributions have been convolved with a Gaussian of width $\sigma=0.17~{\rm mag}$ to represent the width of the RC. In the fiducial model (Fig.~\ref{fig:doublerc}), a bimodality is first evident at $|b| \simeq 5\degrees$ whereas in Model 2 the distribution is only split at $|b| \simeq 6\degrees$.}
\end{figure}

\begin{figure*}
	\centerline{
		\includegraphics[angle=0.,width=.95\hsize]{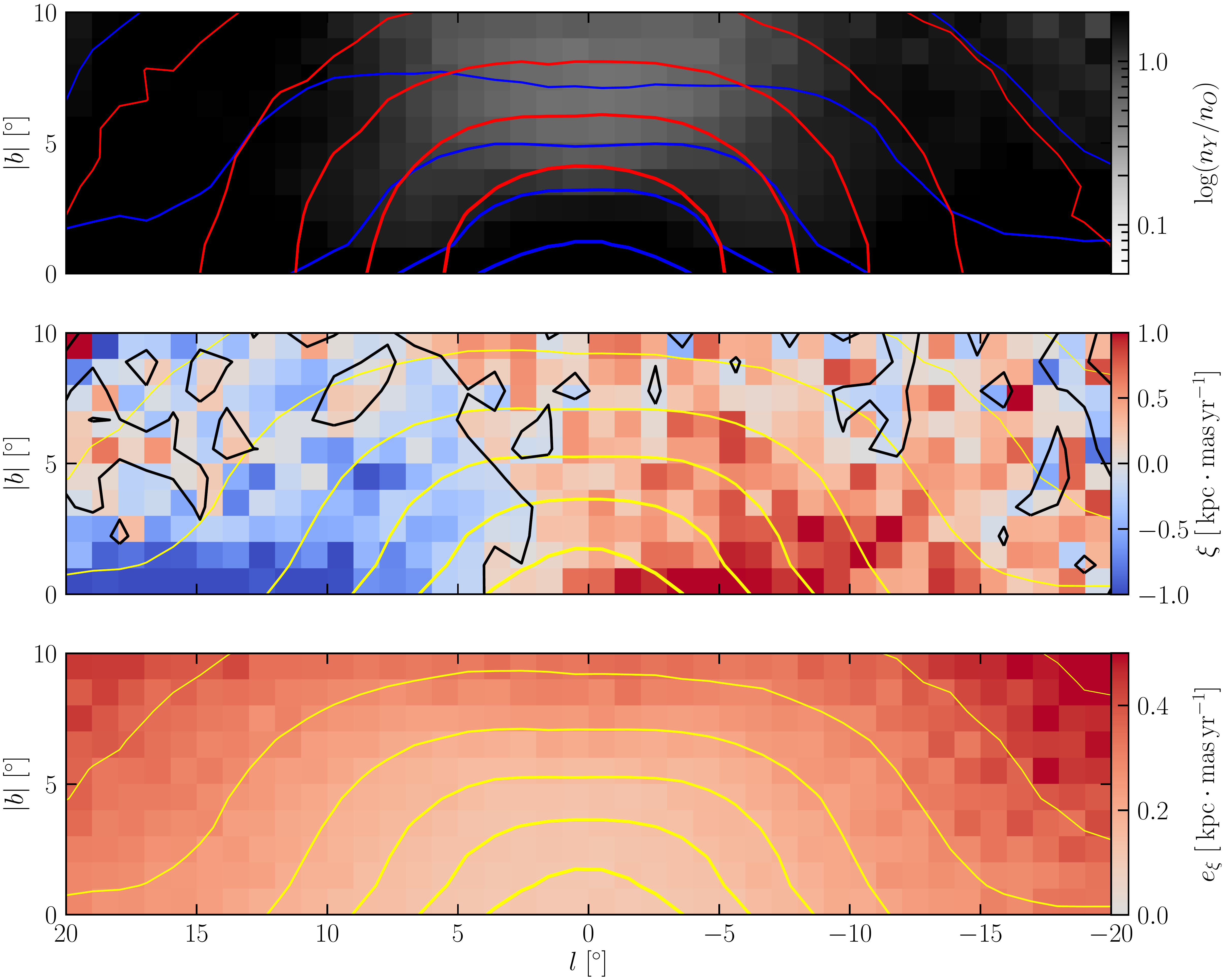}\
	}
	\caption{Top panel: density distribution of bulge stars in Model 2. Blue and red contours follow young and old population densities, respectively. Middle panel: separation amplitude, $\xi$, for each pixel representing a $1\times1~{\rm deg}^2$ field. Bottom panel: model uncertainty on the separation amplitudes for each field. In the bottom two panels, the yellow contours follow the density of all bulge stars.
		\label{fig:field-sepamp-Diff4}}
\end{figure*}

\begin{figure*}
	\centerline{
		\includegraphics[angle=0.,width=.95\hsize]{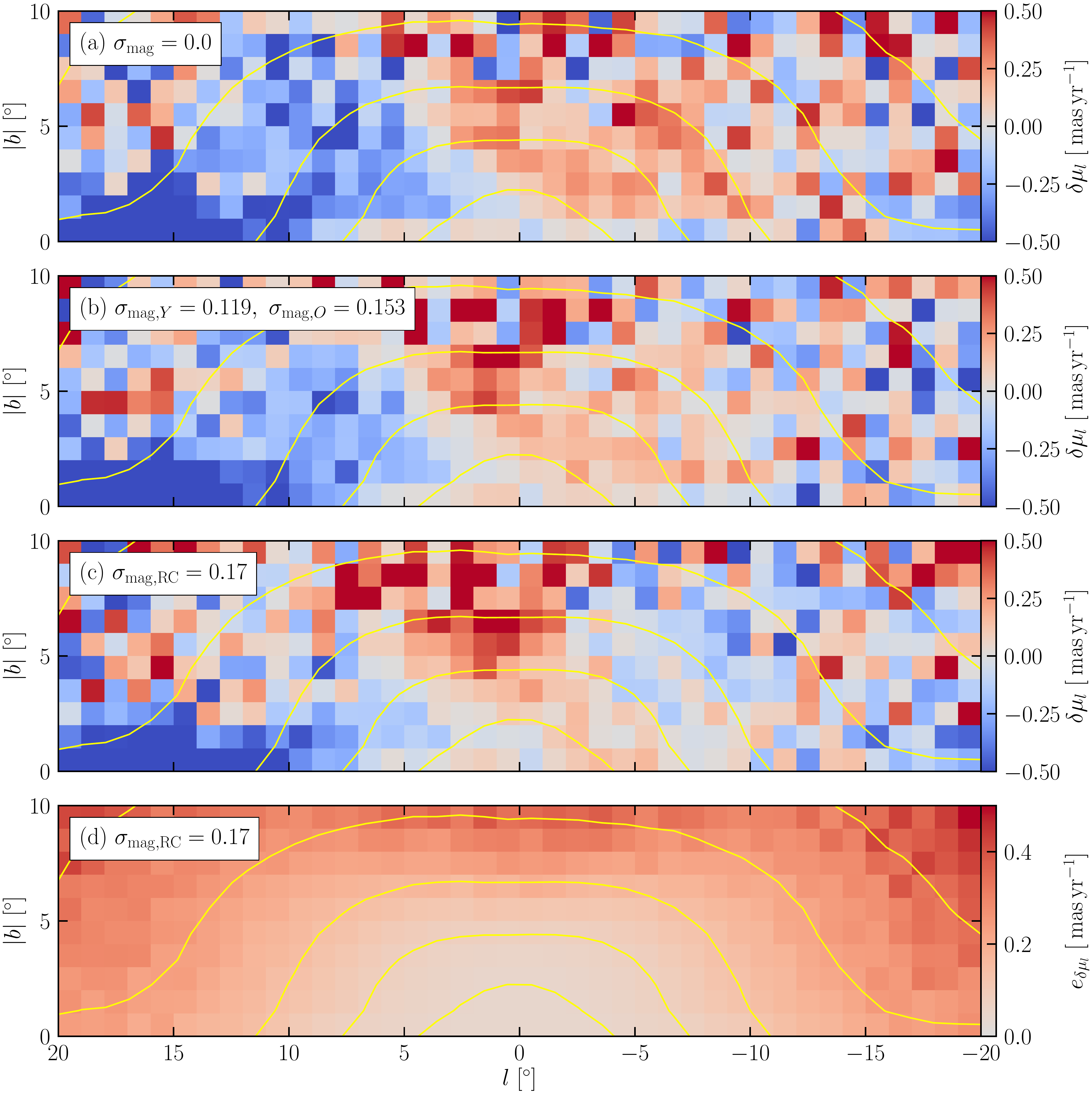}\
	}
	\caption{Top panel: the $\delta \mu_l$ distribution in the bulge region of Model 2 defined as the difference in \apml\ between the young and old populations at $\sim 8 \kpc$. Middle top panel: the same as above but with young and old stars apparent magnitudes convolved with \citetalias{clarkson+18} uncertainties of $\sigma_{\rm mag, Y} = 0.119$ and $\sigma_{\rm mag, O} = 0.153$. Middle bottom panel: the same as above but with both populations convolved with the width of the RC, $\sigma_{\rm mag,RC} = 0.17$. Bottom panel: the calculated error for each field when applying the RC magnitude uncertainties.
		\label{fig:deltav-m2}}
\end{figure*}

%
%
%
%
\bsp	
\label{lastpage}
\end{document}